\PassOptionsToPackage{dvipsnames}{xcolor}

\documentclass{aastex631}
\usepackage{amsmath}
\usepackage{colortbl}

\pdfoutput=1
\usepackage{enumitem}
\usepackage{bm}
\usepackage{multirow}

\usepackage[capitalize]{cleveref}

\usepackage{graphicx}
\usepackage{float}
\usepackage{booktabs}
\usepackage{bigstrut}
\usepackage{capt-of}

\usepackage{physics}
\usepackage{mwe}

\newcommand{\xh}[1]{#1}
\newcommand{\ag}[1]{#1}
\newcommand{\agu}[1]{\textcolor{black}{#1}}
\newcommand{\aguu}[1]{\textcolor{black}{#1}}

\newcommand{\hst}{\emph{HST}\xspace}
\newcommand{\twopr}{\ensuremath{^{\prime \prime}}}
\newcommand{\ho}{\ensuremath{H_0}\xspace}
\newcommand{\emcee}{\texttt{emcee}\xspace}
\newcommand{\lenstr}{\texttt{lenstronomy}\xspace}

\newcommand{\kl}{\normalfont \text{KL}\xspace}
\newcommand{\ess}{\normalfont \text{ESS}\xspace}

\newcommand{\gigalens}{\texttt{GIGA-Lens}\xspace}

\received{February 15, 2022}
\submitjournal{ApJ}

\shorttitle{\gigalens}
\shortauthors{Gu, Huang, et al.}

\begin{document}

\title{\gigalens: Fast Bayesian Inference for Strong Gravitational Lens Modeling}

\author[0000-0003-2748-7333]{A.~Gu}
\affiliation{Department of Physics, University of California, Berkeley, Berkeley, CA 94720}
\affiliation{Department of Electrical Engineering \& Computer Sciences, University of California, Berkeley, Berkeley, CA 94720}

\author[0000-0001-8156-0330]{X.~Huang}
\affiliation{Department of Physics \& Astronomy, University of San Francisco, San Francisco, CA 94117}
\affiliation{Physics Division, Lawrence Berkeley National Laboratory, 1 Cyclotron Road, Berkeley, CA 94720}

\author{W.~Sheu}
\affiliation{Department of Physics, University of California, Berkeley, Berkeley, CA 94720}
\affiliation{Department of Electrical Engineering \& Computer Sciences, University of California, Berkeley, Berkeley, CA 94720}

\author{G.~Aldering}
\affiliation{Physics Division, Lawrence Berkeley National Laboratory, 1 Cyclotron Road, Berkeley, CA 94720}

\author[0000-0001-8156-0330]{A.~S.~Bolton}
\affiliation{NSF's National Optical-Infrared Astronomy Research Laboratory, 950 N. Cherry Ave., Tucson, AZ 85719}

\author[0000-0002-5828-6211]{K.~Boone}
\affiliation{DiRAC Institute, Department of Astronomy, University of Washington, 3910 15th Ave. NE, Seattle, WA 98195}

\author[0000-0002-4928-4003]{A.~Dey}
\affiliation{NSF's National Optical-Infrared Astronomy Research Laboratory, 950 N. Cherry Ave., Tucson, AZ 85719}

\author{A.~Filipp}
\affiliation{Technische Universit\"{a}t M\"{u}nchen, Physik-Department, James-Franck-Stra\ss e 1, 85748 Garching, Germany}
\affiliation{Max-Planck-Institut f\"{u}r Astrophysik, Karl-Schwarzschild-Str. 1, 85748 Garching, Germany}

\author[0000-0002-9253-053X]{E.~Jullo}
\affiliation{Aix-Marseille Univ, CNRS, CNES, LAM, Marseille, France}


\author[0000-0002-4436-4661]{S.~Perlmutter}
\affiliation{Department of Physics, University of California, Berkeley, Berkeley, CA 94720}
\affiliation{Physics Division, Lawrence Berkeley National Laboratory, 1 Cyclotron Road, Berkeley, CA 94720}

\author[0000-0001-5402-4647]{D.~Rubin}
\affiliation{Department of Physics \& Astronomy, University of Hawaii, Honolulu, HI 96822}

\author[0000-0002-3569-7421]{E.~F.~Schlafly}
\affiliation{Lawrence Livermore National Laboratory, 7000 East Avenue, Livermore, CA 94550}

\author[0000-0002-5042-5088]{D.~J.~Schlegel}
\affiliation{Physics Division, Lawrence Berkeley National Laboratory, 1 Cyclotron Road, Berkeley, CA 94720}

\author[0000-0002-9063-698X]{Y.~Shu}
\affiliation{Max-Planck-Institut f\"{u}r Astrophysik, Karl-Schwarzschild-Str. 1, 85748 Garching, Germany}
\affiliation{Ruhr University Bochum, Faculty of Physics and Astronomy, Astronomical Institute (AIRUB), German Centre for Cosmological
Lensing, 44780 Bochum, Germany}


\author[0000-0001-5568-6052]{S.~H.~Suyu}
\affiliation{Technische Universit\"{a}t M\"{u}nchen, Physik-Department, James-Franck-Stra\ss e 1, 85748 Garching, Germany}
\affiliation{Max-Planck-Institut f\"{u}r Astrophysik, Karl-Schwarzschild-Str. 1, 85748 Garching, Germany}
\affiliation{Academia Sinica Institute of Astronomy and Astrophysics (ASIAA), 11F of ASMAB, No.1, Section 4, Roosevelt Road, Taipei
10617, Taiwan}

\correspondingauthor{Xiaosheng Huang, Andi Gu}
\email{xhuang22@usfca.edu, andi.gu@berkeley.edu}


\begin{abstract}
We present \gigalens: a gradient-informed, GPU-accelerated Bayesian framework for modeling strong gravitational lensing systems, implemented in TensorFlow and JAX.  The three components, optimization using multi-start gradient descent, posterior covariance estimation with variational inference, and 
sampling via Hamiltonian Monte Carlo\agu{,} all take advantage of gradient information through automatic differentiation and massive parallelization on graphics processing units (GPUs). 
We test our pipeline on a large set of simulated systems and demonstrate in detail its high level of performance.  
The average time to model a single system on four Nvidia A100 GPUs is 105 seconds.  
The 
robustness, speed, and scalability offered by this framework make it possible to model the large number of strong lenses found in current surveys and \agu{present a very} promising prospect for the modeling of 
$\mathcal{O} (10^5)$ lensing systems expected to be discovered in the era of the \agu{Vera C.} Rubin Observatory, Euclid, and the \agu{Nancy Grace} Roman Space Telescope.

\end{abstract}

\keywords{cosmology -- gravitational lensing: strong -- methods: statistical}

\section{Introduction}
\label{sec:intro}
Strong gravitational lensing systems are a powerful tool for cosmology. They have been used to study how dark matter is distributed in galaxies and clusters \citep[e.g.,][]{kochanek1991a, hogg1994, broadhurst2000a, koopmans2002a, bolton2006a,bradac2008a,koopmans2006a, vegetti2009a,  huang2009a, jullo2010a, grillo2015a,shu2015a,shu2016b,shu2017, meneghetti2020a},  
and are uniquely suited to probe the low-end of the dark matter mass function and test the prediction of the cold dark matter (CDM) model beyond the local universe \citep[e.g.,][]{vegetti2010a, vegetti2012a, hezaveh2016a, ritondale2019a, diazrivero2020a, cagansengul2020a,sengul2021,gilman2021}. 
Multiply lensed supernovae (SNe) are ideal for measuring time delays and \ho because of their well-characterized light curves, and in the case of Type~Ia, with the added benefit of standardizable luminosity \citep{refsdal1964a, treu2010a, oguri2010a}\agu{, provided microlensing can be accurately characterised \citep{yahalomi2017,foster2018}. Furthermore, SNe have the benefit of fading, so for these systems, lens models can be validated using images that are uncontaminated by bright point sources \citep{ding2021}.}
In recent years, strongly lensed supernovae,  both core-collapse \citep{kelly2015a, rodney2016a} and Type~Ia \citep{quimby2014a, goobar2017a, rodney2021a}, have been discovered. 
Time-delay \ho measurements from multiply imaged supernovae \citep[e.g.,][]{goldstein2017a, shu2018, goldstein2018b, goldstein2018a, pierel2019a, suyu2020a, huber2021a}, 
combined with measurements from distance ladders \citep[e.g.,][]{riess2019, freedman2019a, freedman2020a,riess2021a} and lensed quasars \citep[e.g.,][]{suyu2010a, suyu2013a, treu2016a, bonvin2017a, wong2019a,millon2020a,birrer2020a}, can be an important test of the tension  between \ho measured locally and the value inferred from the \agu{Cosmic Microwave Background \citep[CMB;][]{planck2020}}.

The introduction of neural networks to identify gravitational lens candidates in imaging surveys has 
been transformational \citep[e.g.,][]{jacobs2017a, metcalf2018a, jacobs2019a, jacobs2019b, canameras2020a}. 
In our recent work, 
we discovered over 1500  
new strong lenses \citep{huang2020a, huang2021a}
in the Dark Energy Spectroscopic Instrument (DESI) Legacy Imaging Surveys \aguu{\citep{dey2019}} by using residual neural networks. 
This trend will accelerate in the era of the \agu{Vera C. Rubin Observatory} Legacy Survey of Space and Time (LSST), Euclid, and the \agu{Nancy Grace} Roman Space Telescope, 
when $\mathcal{O}(10^5)$ strong lensing systems are projected to be found in the next decade \citep{collett2015a}. 
\xh{As} the lens search efficiency has dramatically improved, 
\xh{there has been significant development on strong lens modeling as well.
For example, the widely-used
\lenstr \citep{birrer2018a} is a fully-fledged (including, for example, a suite of plotting routines) and highly versatile lens modeling package.}
However, the computational cost for lens modeling remains high.
For instance, \citet{rojas2021a} \xh{modeled 
41 systems with a single deflector in $gri$ bands from the Dark Energy Survey using a pipeline based on \lenstr. 
They used S\'ersic profiles for lens and source light, and for the lensing potential, a singular isothermal ellipsoid model  and external shear.
They reported that modeling a single system took 4.3~hours on average.}
In addition, \lenstr uses particle swarm optimization (PSO). PSO is an easily parallelizable heuristic algorithm for non-convex optimization. However, it does not have any guarantee of convergence, especially in high dimensions (see \S\ref{sec:map}).
Furthermore, 
\xh{\emcee \citep{foreman2013a}, a popular MCMC algorithm in astrophysics, including being used in \lenstr,}
relies on sampling techniques that also show undesirable behavior in high dimensional parameter spaces (see \S\ref{sec:hmcvemcee}).
For lens modeling using high resolution images, often more complex lens and source models
will be used, and this can significantly increase the dimensionality of the parameter space.
\xh{For strong gravitational lenses to realize their full potential as an effective probe for cosmology,
it is crucial that we address these issues and 
make the process of modeling lensing systems 
\xh{robust, considerably faster, and scalable to high-dimensional parameter spaces}.} 

In this paper, we present a Bayesian lens modeling framework that fulfills these requirements. 
We describe our \agu{gradient-informed, GPU-accelerated (\gigalens) framework} in \S\ref{sec:modeling}.
In \S\ref{sec:results}, we 
\xh{demonstrate the performance of} our pipeline on a large set of simulated systems. 
We discuss our results and conclude in \S\ref{sec:conclusion}.

\section{Lens Modeling}
\label{sec:modeling}
In this section, we introduce \agu{the \gigalens\footnote{\agu{Our framework will be made public upon acceptance of this paper.}}} modeling framework. 
In the initial stages of the development of our pipeline, \lenstr served as a helpful guide,
specifically\agu{, in its approach of using optimization to find a region of high posterior density from which the MC sampler can be initialized}. We believe our framework represents a significant improvement upon the \texttt{lenstronomy} modeling pipeline \agu{in terms of speed, optimization, and sampling}.
Our entire framework is  
implementated in both TensorFlow \citep{TensorFlow} and JAX \citep{bradbury2018a}. Complete integration with either of these libraries confers significant advantages.
It enables seamless execution of our code on graphics processing units (GPUs), which 
can achieve much faster gravitational lens simulation and modeling. 
Even modest, freely available GPUs
are capable of performing basic linear algebra operations 
(which are at the core of lens modeling codes) one to two orders of magnitude faster than a typical CPU. 
In addition, our tight integration with TensorFlow allows us to use the TensorFlow Probability \citep{dillon2017a} library for probabilistic modeling (which also provides support for JAX). \ag{Conveniently, this library} has already implemented 
advanced statistical methods such as variational inference and adaptation algorithms (for step size and trajectory length; \citet{hoffman2014,hoffman2021}) in Hamiltonian Monte Carlo. 
\ag{These features play a central role in our pipeline.}
In \S\ref{sec:model-spec}, we describe our lens model, both its physical and probabilistic aspects.
Obtaining the gradient for the posterior with automatic differentiation is presented in \S\ref{sec:ad}. 
Next, we detail the three main steps in our modeling pipeline.
In \S\ref{sec:map}, we find the global optimal values for the lensing parameters using multi-start gradient descent.  
In \S\ref{sec:vi}, we estimate the covariance matrix around the global optimum using variational inference.  
This allows us to sample the posterior distribution efficiently with Hamiltonian Monte Carlo in \S\ref{sec:hmc}.

\subsection{Model Specification}\label{sec:model-spec}

\subsubsection{Physical Model}\label{sec:phys-model}
\begin{figure}
    \centering
    \includegraphics[width=0.65\textwidth]{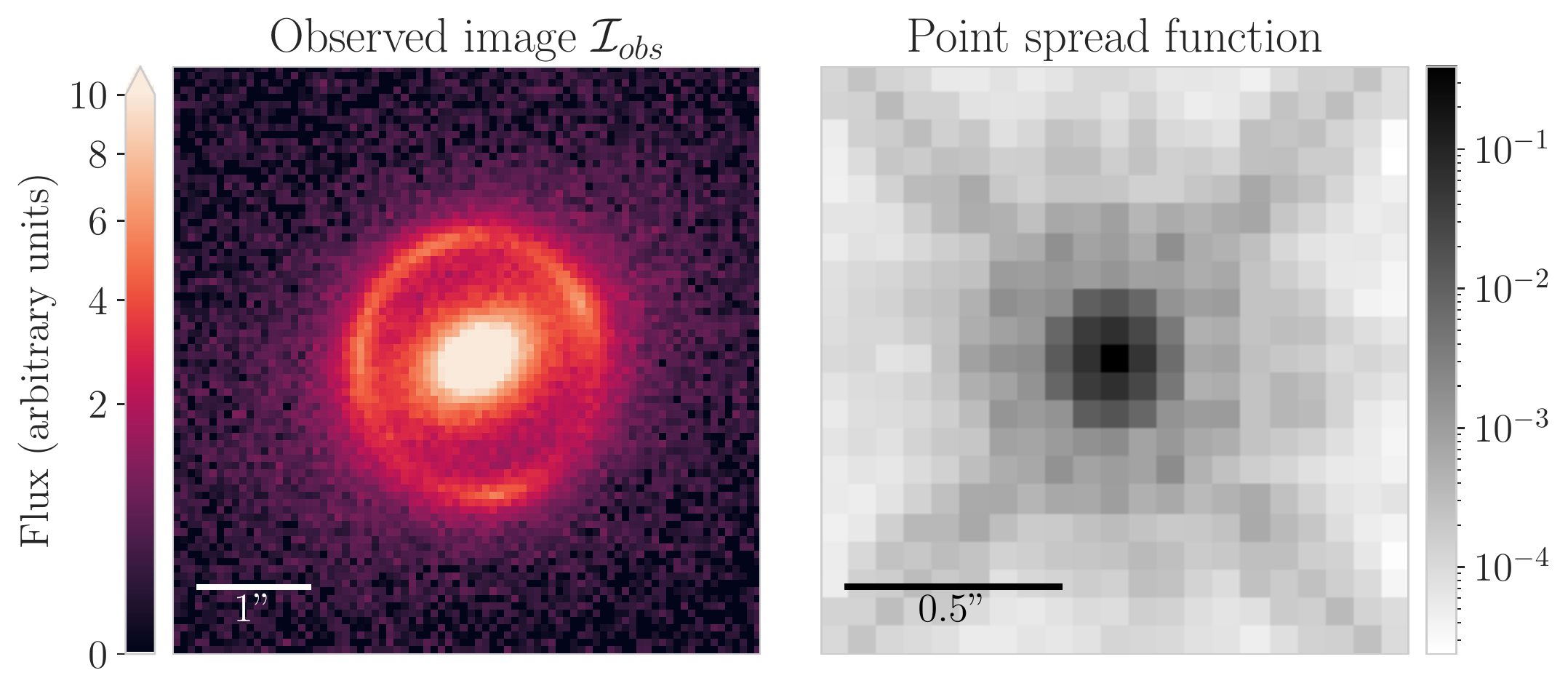}
    \caption{A lensing system simulated using \texttt{lenstronomy} is shown on the left. 
    The parameters for this system are taken directly from the \texttt{lenstronomy} starting guide Jupyter Notebook \citep{birrer2021}.
    Although \texttt{lenstronomy} uses a Gaussian point spread function (PSF) for this system, we use a more realistic PSF calculated for the Hubble Space Telesopce (\hst) WFC3 F140W band using \texttt{TinyTim} \citep{krist2011a} with \agu{an original pixel scale of 0.13\twopr. We then subsample this PSF to a scale of 0.065\twopr, as shown on the right.} This system will be used to show the step-by-step workflow of our modeling pipeline. It will henceforth be called the reference system.}
    \label{fig:ref-sys}
\end{figure}

\agu{Given a lens model fully described by a set of parameters $\Theta$, the predicted \aguu{counts per second} at arbitrary positions on the image plane, $\mathcal{I}_{model}(x,y; \Theta)$, can be determined by evaluating the deflection angle at $(x,y)$, using this deflection to ray trace \citep[e.g.,][]{narayan1997} onto the source plane, \aguu{setting the surface brightness of the lensed source at $(x,y)$ to the corresponding surface brightness on the source plane (via surface brightness conservation),} then adding the lens light, and finally convolving with the PSF\footnote{\agu{The sky brightness, $I_{sky}$, can also be included in the model parameters $\Theta$. In this work, we assume the sky brightness has been subtracted.}}. To demonstrate how our modeling pipeline works, we simulate a reference system (\cref{fig:ref-sys}) using \lenstr. The lens model used for this simulation consists of an elliptical power law\footnote{We note the EPL model is equivalent to the power law elliptical mass density profile \citep[PEMD;][]{barkana1998a}.} (EPL) mass model \aguu{for the lens} \citep{tessore2015a} and external shear.} The EPL model is characterized by the surface mass density \ag{in units of the critical density}, or convergence,
\begin{equation}
    \kappa(x_{\agu{lens}},y_{\agu{lens}}) = \frac{3-\gamma_{epl}}{2} \qty(\frac{\theta_E}{\sqrt{qx_{\agu{lens}}^2+y_{\agu{lens}}^2/q}})^{\gamma_{\agu{epl}}-1},
\end{equation}
where $\theta_E$ is the Einstein radius, $\gamma_{epl}$ is the mass profile slope, 
the coordinates $x_{\agu{lens}},y_{\agu{lens}}$ are aligned with the major and minor axes of the lens, and $q$ is the axial ratio. \agu{The transformation between the image coordinates $(x,y)$ and the lens\agu{-centric} coordinates $(x_{lens},y_{lens})$ is \begin{equation}
     \mqty[x_{lens} \\ y_{lens}] = \mqty[\cos \phi & \sin \phi \\ -\sin \phi & \cos \phi] \mqty[x-x_{epl} \\ y-y_{epl}] \label{eqn:coord-trans}
\end{equation}
where $\phi$ is the position angle and $x_{epl}$ and $y_{epl}$ are the lens center.}
The deflection angle for this model can be written in terms of the Gaussian hypergeometric function, which can be calculated iteratively, converging to a high degree of accuracy in \aguu{a small number} of iterations. The truncation error reaches $\lesssim 10^{-16}$ within 35 iterations for $q \gtrsim 0.5$ \citep{tessore2015a}. 


To include the effects of the local environment, we include an external shear with a deflection angle
\begin{equation}
    \alpha_{ext}(x,y) = \qty(\gamma_{ext,1} x + \gamma_{ext,2} y, \gamma_{ext,2} x - \gamma_{ext,1} y).
\end{equation}
Finally, for our simulations, we model our lens and source light with an elliptical S\'ersic profile \citep{sersic1963}:
\begin{equation}
    I(x_{\agu{light}},y_{\agu{light}}) = I_0 \exp(-b_n \qty(\qty(\frac{\sqrt{qx_{\agu{light}}^2 + y_{\agu{light}}^2/q}}{R_{eff}})^{1/n} - 1)),
\end{equation}
where $b_n=1.9992n - 0.3271$, $R_{eff}$ is the \agu{effective radius (half-light radius)}, $n$ is the S\'ersic index, and $x_{\agu{light}},y_{\agu{light}}$ are 
aligned with the major and minor axes of the light profile. \agu{The transformation between image coordinates $(x,y)$ and light coordinates $(x_{light}, y_{light})$ is identical in form to \cref{eqn:coord-trans}. We will use $(x_l, y_l)$ to denote the lens light center \aguu{(in this work, the lens light center is not fixed to the lens mass center)} and $(x_s,y_s)$ to denote the source light center. Finally, in lens modeling, eccentricities are often used through the reparameterization,
\begin{equation}
    (\epsilon_1,\epsilon_2) = \frac{1-q}{1+q} \qty(\cos(2\phi), \sin(2\phi)),
\end{equation}
where $\phi$ is \agu{the} position angle. We will use $(\epsilon_{epl,1},\epsilon_{epl,2})$ to denote the lens mass eccentricities, $(\epsilon_{l,1},\epsilon_{l,2})$ for the lens light eccentricities, and $(\epsilon_{s,1},\epsilon_{s,2})$ for the source light eccentricities.} \aguu{Similarly, $R_l$ and $R_s$ will denote the lens and source light effective radius.}

\agu{In practice, $\mathcal{I}_{model}(x,y; \Theta)$ is vectorized to be evaluated on a grid of pixels simultaneously. This grid is supersampled by some integer factor $k_{super}$ (in this work, $k_{super}=2$) from the image coordinates, so there is an additional step in evaluating $\mathcal{I}_{model}$ that consists of downsampling (by averaging) the predicted image by $k_{super}$ back to the image coordinate grid.}


\subsubsection{Probabilistic Model}\label{sec:prob-model}

Our probabilistic model 
comprises a likelihood function $\mathcal{L}(\Theta; \mathcal{I}_{obs}) \equiv p(\mathcal{I}_{obs} \mid \Theta)$ and a prior $p(\Theta)$, where $\Theta \in \mathbb{R}^d$ are the lensing system parameters, and $\mathcal{I}_{obs}$ is the observed image (in units of counts/sec). 
The likelihood function requires a preprocessed
observed image $\mathcal{I}_{obs}$ as well as specification of the background (sky) Gaussian noise $\sigma_{bkg}$ and exposure time $t_{exp}$. For this work, we define:
\begin{equation}\label{eqn:model-def}
\begin{gathered}
    \log \mathcal{L}(\Theta; \mathcal{I}\agu{_{obs}}) = -\frac{1}{2} \sum_{x,y} \qty[ \frac{\qty(\mathcal{I}_{obs}(x,y)-\mathcal{I}_{model}(x,y; \Theta))^2}{\sigma_{tot}^2(x,y; \Theta)} + \log\qty(2\pi\sigma_{tot}^2(x,y; \Theta))] \\
    \sigma_{tot}^2(x,y; \Theta) = \sigma_{bkg}^2 + \frac{\mathcal{I}_{model}(x,y; \Theta)}{\mathcal{G} \cdot t_{exp}},
\end{gathered}
\end{equation}
where $\mathcal{I}_{model}(x,y; \Theta)$ is our \aguu{forward} model for the observed image, $\mathcal{G}$ is the gain \aguu{(in $\mathrm{e^-}$/count)}, and the sum is over all pixels $(x,y)$ in the observed data. 
\agu{The second term in the variance map arises from Poisson shot noise: since the \aguu{expected value of the electron counts is $t_{exp} \mathcal{I}_{model} \mathcal{G}$, the Poisson variance is therefore also $t_{exp} \mathcal{I}_{model} \mathcal{G}$, and so the Poisson variance of the model image is $\frac{t_{exp} \mathcal{I}_{model} \mathcal{G}}{\mathcal{G}^2 \cdot t_{exp}^2} = \frac{\mathcal{I}_{model}}{\mathcal{G} \cdot t_{exp}}$}. The form of this likelihood assumes independent per-pixel noise. However, the variance map $\sigma_{tot}^2$ can be specified to account for correlated pixels and to incorporate prior information about the noise at each pixel. For example, an alternative form for $\sigma_{tot}^2$, defined in terms of the observed image, is sometimes used: $\sigma_{tot}^2(x,y) = \sigma^2_{bkg} + \frac{\mathcal{I}_{obs}(x,y)}{ \mathcal{G} \cdot t_{exp}}$ , so that the total variance is independent of $\Theta$.}
This is done for computational efficiency.
For example, it allows for linear light profile parameters to be solved for in closed form. \agu{We note, however, this definition for the variance may induce biases at low signal-to-noise ratios \citep{horne1986}.\footnote{\ag{Another issue with using $\mathcal{I}_{obs}$ to define the noise map $\sigma^2_{tot}$ is that it may have negative pixel values due to Gaussian noise. In \lenstr, the negative pixels are set to zero, which is not the most rigorous way to estimate the Poisson noise. Despite the slight difference in these two approaches, 
for our simulated systems (\S\ref{sec:results}), the results using both are virtually identical.}} Therefore}, we opt to use the more rigorous definition.
In our pipeline, this incurs little additional computational cost.
\agu{We also point out that the first term in the log-likelihood corresponds to \begin{equation}
    \chi^2 = \sum_{x,y} \frac{\qty(\mathcal{I}_{obs}(x,y)-\mathcal{I}_{model}(x,y; \Theta))^2}{\sigma_{tot}^2(x,y; \Theta)},
\end{equation}
and the second term is the normalization factor.}


The prior for the parameters is typically defined as a product of independent distributions, \agu{each of which} can be tuned either to reflect 
prior knowledge about lensing parameters, or to match a physical understanding of the particular system that is being modelled (e.g., the Einstein radius can typically be estimated to within 20\% from visual inspection).
Given our physical understanding of the model, our prior ought to vanish for certain regions of parameter space, such as $\theta_E<0$. 
In our modeling, we make use of mappings that naturally enforce these constraints. We \agu{describe these mappings} in \S\ref{sec:ad}.

\newcommand{\scale}[2]{\textcolor{red}{#1} \ / \ \textcolor{blue}{#2}}
\agu{In this work, we use a ``simulation distribution" to generate \aguu{a full set of 22 parameters for} 100 lensing systems, and the prior distribution used to model these lenses (see \S\ref{sec:results}) is a broadened version of this simulation distribution \aguu{(note some rows contain two parameters)}:}
\begin{equation}
\begin{split}\label{eqn:sim-dist}
\text{Lens mass}: &
\left\lbrace
\array{@{}r@{\quad}l@{}r@{}}%
\theta_{E} &\sim \exp(\mathcal{N}(\ln 1.25, \scale{0.25}{0.4})) & \hspace{78 pt} \text{[Einstein radius(\twopr)]} \\
\gamma_{epl} &\sim \mathcal{TN}(2, \scale{0.25}{0.5}; 1, 3) & \text{[Mass slope]} \\
\epsilon_{epl,1}, \epsilon_{epl,2} &\sim \mathcal{N}(0, \scale{0.1}{0.2}) & \text{\qquad [Lens mass eccentricities]} \\
x_{epl}, y_{epl} &\sim \mathcal{N}(0, \scale{0.03}{0.06}) & \text{[Lens mass center(\twopr)]} \\
\gamma_{ext,1}, \gamma_{ext,2} &\sim \mathcal{N}(0, \scale{0.05}{0.1}) & \text{[External shear components]}
\endarray \right. \\
\text{Lens light}: &\left\lbrace
\array{@{}r@{\quad}l@{}r@{}}%
R_{l} &\sim \exp(\mathcal{N}(\ln 1.6, \scale{0.15}{0.25})) & \hspace{65 pt}  \text{[Lens S\'ersic radius(\twopr)]} \\
n_{l} &\sim \mathcal{U}(\scale{2}{0.5}, \scale{6}{8}) & \text{[Lens S\'ersic index]} \\
\thinspace \epsilon_{l,1}, \epsilon_{l,2} &\sim \mathcal{TN}(0, \scale{0.05}{0.1}; -0.15, 0.15) & \text{[Lens light eccentricities]}\\
x_{l}, y_{l} &\sim \mathcal{N}(0, \scale{0.01}{0.02}) & \text{[Lens light center(\twopr)]} \\
I_l &\sim \exp(\mathcal{N}(\ln 300, \scale{0.3}{0.5})) & \text{[Lens half light intensity]}
\endarray \right. \\
\text{Source light}: &\left\lbrace
\array{@{}r@{\quad}l@{}r@{}}%
R_s &\sim \exp(\mathcal{N}(\ln 0.25, \scale{0.15}{0.25})) & \hspace{65 pt} \text{[Source S\'ersic radius(\twopr)]} \\
n_s &\sim \mathcal{U}(0.5, \scale{4}{8}) & \text{[Source S\'ersic index]} \\
\thinspace \epsilon_{s,1}, \epsilon_{s,2} &\sim \mathcal{TN}(0, \scale{0.15}{0.3}; -0.5, 0.5) & \text{[Source light eccentricities]}\\
x_s, y_s &\sim \mathcal{N}(0, \scale{0.25}{0.5}) & \text{[Source light center(\twopr)]} \\
I_s &\sim \exp(\mathcal{N}(\ln 150, \scale{0.5}{0.9})) & \text{[Source half light intensity]}
\endarray \right.
\end{split}
\end{equation}
where $\mathcal{U}(a,b)$ is a uniform distribution with support $\qty[a,b]$, $\mathcal{N}(\mu, \sigma)$ is Gaussian with mean $\mu$ and standard deviation $\sigma$, and $\mathcal{TN}(\mu, \sigma; x_{low}, x_{high})$ is a truncated Gaussian with support $\qty[x_{low}, x_{high}]$. \agu{For the distribution parameters, we use the notation $\scale{a}{b}$ to indicate that the simulation distribution uses the parameter $\textcolor{red}{a}$ while the prior uses $\textcolor{blue}{b}$. For instance, when generating our dataset, we sample $x_{epl}$ from $\mathcal{N}(0, 0.03)$, while during modeling, our prior for $x_{epl}$ is $\mathcal{N}(0, 0.06)$.}

\begin{figure}[H]
    \centering
    \includegraphics[width=\textwidth]{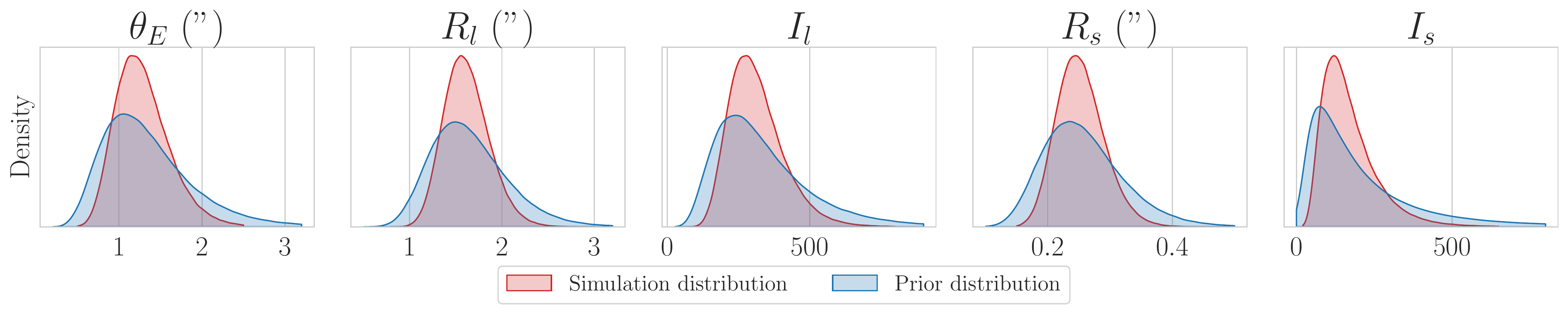}
    \caption{Probability density for five components of the simulation and prior distribution that are log-normal.}
    \label{fig:prior}
\end{figure}

Note that $\exp(\mathcal{N}(\ldots))$ is a log-normal distribution, \ag{used for 5 parameters. Since these distributions are not as intuitive, we show them in \cref{fig:prior}}. 
The simulation distribution for the light amplitudes $I_l$ and $I_s$ have been chosen in such a way that the typical signal-to-noise ratio of the arcs is 100 (with a range between 30 and 200), and is comparable to the amplitudes used in the \lenstr starting guide Jupyter Notebook \citep{birrer2021}.

\begin{figure}[H]
    \centering
    \includegraphics[width=0.6\textwidth]{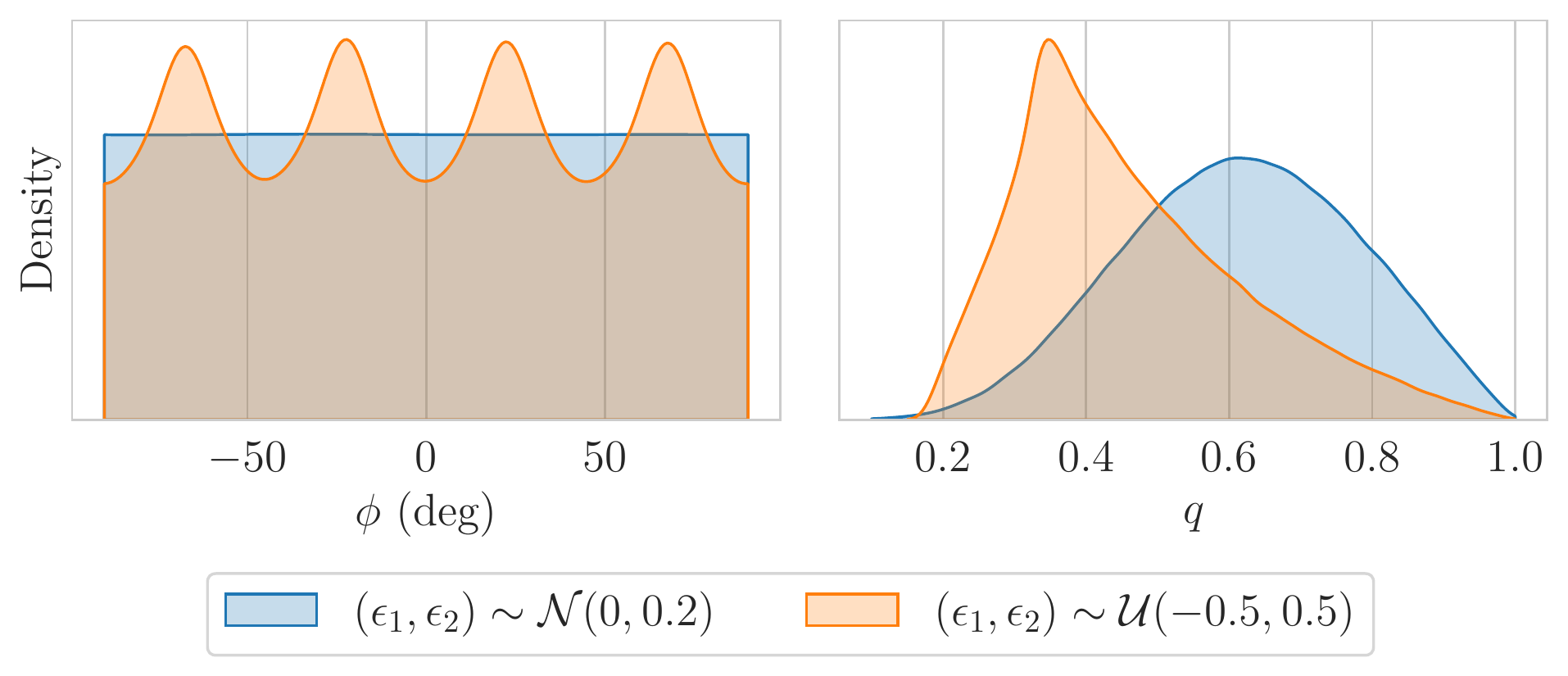}
    \caption{The corresponding distributions for $\phi, q$ when the ellipticities are normally distributed (blue) and uniformly distributed (orange).}
    \label{fig:prior-choice}
\end{figure}

\agu{Although the distributions in \cref{eqn:sim-dist} can be specified in any format, in this work, we make a number of deliberate choices. Most importantly, for parameters such as ellipticities (e.g., $\epsilon_{epl,1},\epsilon_{epl,2}$) and centers (e.g., $x_{epl},y_{epl}$), we use normal distributions to reflect their rotational symmetry. For instance, in \cref{fig:prior-choice}, we show how a normal distribution for the ellipticities results in a uniform distribution for the position angle $\phi$, whereas independent uniform distributions for the ellipticities break rotational symmetry, resulting in a non-uniform distribution for $\phi$ and a highly non-Gaussian distributions for $q$.}

\subsection{Obtaining Gradient Information}\label{sec:ad}
For the models discussed in \S\ref{sec:phys-model}, $\mathcal{I}_{model}(x, y; \Theta)$ is a differentiable mapping with respect to the model parameters, $\Theta$. Therefore, so is the posterior density \ag{(\cref{eqn:model-def})} within the support of the prior\footnote{\agu{Although the gradient is continuous everywhere, it may vanish in certain regions of the parameter space due to extreme misalignment of the lens and source. We avoid this by setting our prior in a way such that, $\sqrt{(x_{s}-x_{epl})^2+(y_s-y_{epl})^2} \sim \theta_E$.}}. This is essential. As we will show in \S\ref{sec:map} and \S\ref{sec:hmc}, gradient information about the posterior is 
necessary to successfully achieve fast convergence for complex lens models. Although the posterior is theoretically differentiable, to obtain gradient with numerical differentiation (i.e., finite difference) would be prohibitively expensive.
Therefore, we exploit the fact that \aguu{the mapping} $\mathcal{I}_{model}$ is composed of a sequence of differentiable operations (e.g., \ag{convolution}) and use automatic differentiation \citep[AD;][]{wengert1964}. 
Crucially with AD\ag{, which is implemented in both TensorFlow and JAX}, the \agu{additional} computational cost of \agu{evaluating} the gradient is independent of the number of parameters \citep{baydin2018}. The full gradient $\pdv{\mathcal{I}_{model}}{\Theta_i}$ that we are interested in can be calculated in approximately the same amount of time it takes to calculate $\mathcal{I}_{model}$ itself.

Although we can calculate the gradient efficiently with AD, there are regions of parameter space where this gradient is undefined as a result of having hard boundaries for the parameters. This can have a number of undesirable effects for modeling. 
For example, any algorithm with an iterative update that has a finite step size for the parameters may result in certain solutions being updated to positions outside the support of the prior. 
Therefore we instead use smooth, invertible functions (termed ``bijectors''), $g$, to map from an unconstrained space to the support of the prior, $\text{supp}(p)$. 
For instance, the Einstein radius has a semi-infinite support $\mathbb{R}_{>0}$, so we use the exponential map
\begin{equation}
\begin{split}\label{eqn:exp}
    g_{\theta_E}: 
    z &\mapsto e^z
\end{split}
\end{equation}
For other parameters, such as the mass profile slope $\gamma$, that may have a finite support $\qty(a,b)=(1,3)$, 
a convenient bijector to use is the sigmoid mapping:
\begin{equation}\label{eqn:sigmoid}
\begin{split}
    g_\gamma: 
    z &\mapsto a + \frac{b-a}{1+e^{-z}}
\end{split}
\end{equation}
The joint bijector $g: \mathbb{R}^d \rightarrow \text{supp}(p)$ is typically constructed component-wise from the bijectors for each parameter.\footnote{\agu{Although $g$ can be specified in any desired form, we find that simply constructing it component-wise from the standard bijectors (\cref{eqn:exp,eqn:sigmoid}) is sufficient. This component-wise construction is also inexpensive to evaluate, since each of the standard bijectors are elementary.}} That is, $\Theta=g(\tilde{\Theta})$, 
using the notation $\tilde{\cdot}$ to henceforth denote quantities in the unconstrained space.
This allows us to optimize or sample over the unconstrained space $\mathbb{R}^d$, rather than manually enforcing constraints on the parameters. 
Since the volume element in the unconstrained space is different from the constrained parameter space, the prior needs to be modified,
\begin{equation}
    \log \tilde{p}(\tilde{\Theta}) = \log p(g(\tilde{\Theta})) + \log \abs{J},
\end{equation}
where $J$ is the Jacobian of $g$ evaluated at \agu{$\tilde{\Theta}$}.
The likelihood remains unchanged,
\begin{equation}
    \log \tilde{p}(\mathcal{I}_{obs} \mid \tilde{\Theta}) = \log p(\mathcal{I}_{obs} \mid g(\tilde{\Theta})).
\end{equation}
Thus for the posterior,
\begin{equation}
    \log \tilde{p}(\tilde{\Theta} \mid \mathcal{I}_{obs}) = \log p(\Theta \mid \mathcal{I}_{obs}) + \log \abs{J}. \label{eqn:reparam-post}
\end{equation}
The constrained parameters that correspond to physical quantities in the lens model (e.g., $\theta_E$) are hereafter called physical parameters.
With the posterior distribution thus reparameterized, it is straightforward to compute its derivatives using AD.

\subsection{Maximum a Posteriori Estimate}\label{sec:map}
When sampling from a high dimensional posterior using Monte Carlo (MC) algorithms, since all but a small fraction of the parameter space has a vanishing posterior density, arbitrarily chosen initializations for these samplers will take a long time to converge to the posterior distribution. 
Therefore, it is necessary to identify a region of high posterior density from which MC samplers can be initialized. Typically, the maximum a posteriori (MAP) estimate 
serves this purpose well:
\begin{equation}
    \tilde{\Theta}^*_{MAP} = \underset{\tilde{\Theta} \in \mathbb{R}^d}{\text{argmax}} \ \log \tilde{p}(\tilde{\Theta} \mid \mathcal{I}_{obs}) = \underset{\tilde{\Theta} \in \mathbb{R}^d}{\text{argmin}} \ \qty(-\log \tilde{p}(\mathcal{I}_{obs} \mid \tilde{\Theta}) - \log \tilde{p}(\tilde{\Theta})).
\end{equation}
Here and below, the notation $\cdot^*$ will denote a local optimum, 
and the notation $\cdot^*_{MAP}$ will indicate that this local optimum is also globally optimal. Finding $\tilde{\Theta}_{MAP}^*$ is an optimization task with the objective function $f$ being the negative (unnormalized) log posterior density:
\begin{equation}
f(\tilde{\Theta})= -\qty(\log \tilde{p}(\mathcal{I}_{obs} \mid \tilde{\Theta}) + \log \tilde{p}(\tilde{\Theta})). \label{eqn:obj-fun}
\end{equation}

\begin{figure}[h!]
\begin{minipage}{0.56\textwidth}
\includegraphics[width=\textwidth]{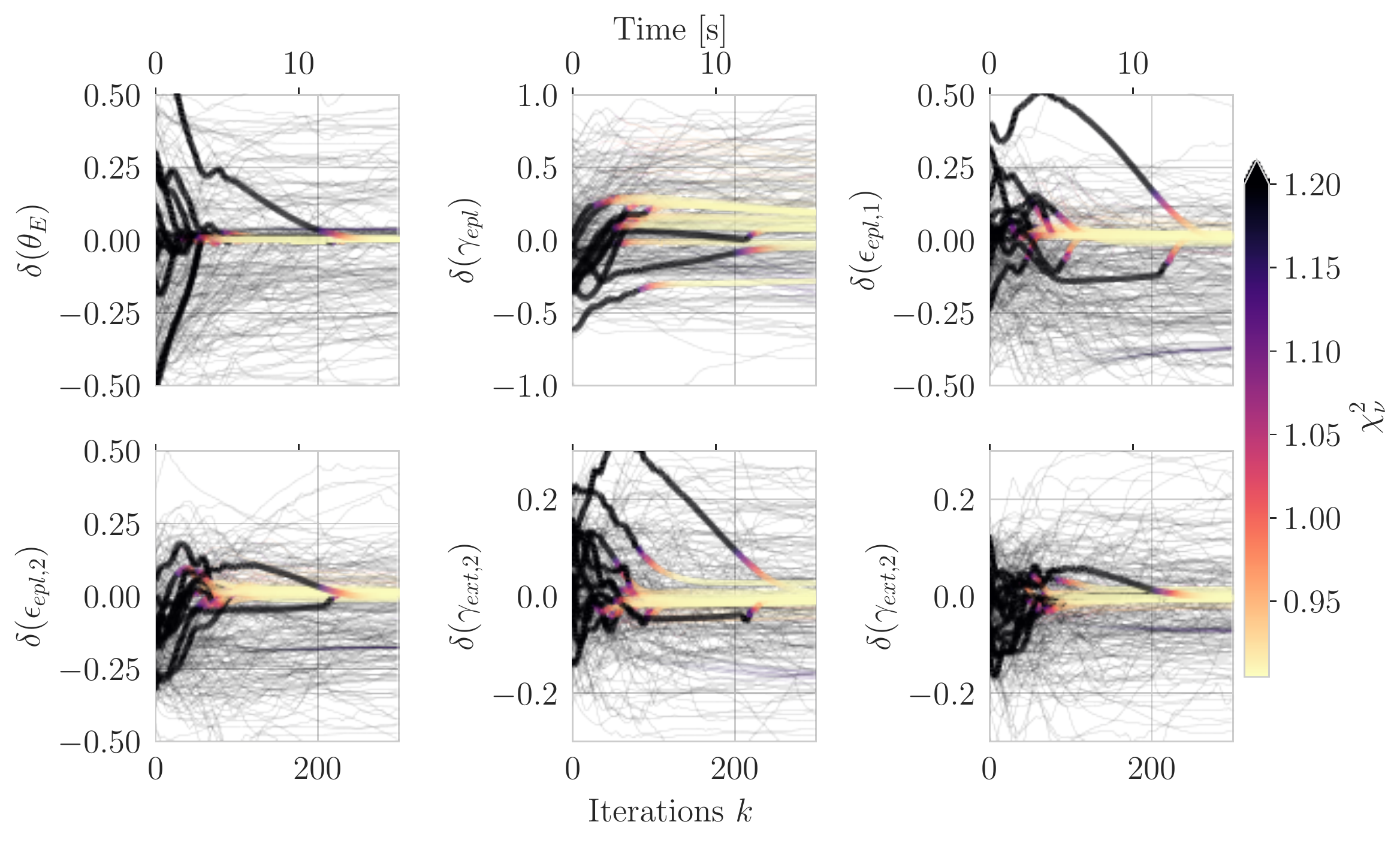}
\caption{The error trajectories of the lensing parameters (i.e., the difference between the $k$th iterate and the \agu{ground truth}) for the reference system (see \cref{fig:ref-sys}) over the course of gradient descent. Trajectories that end in $\chi_\nu^2 \leq 1.01$ \agu{(where $\chi_\nu^2=\chi^2/\text{DOF}$)} are shown with thicker lines\aguu{, all of which converge to the global minimum (see \cref{fig:ref-loss})}. We note that a wide range of samples converge to the correct solution (see text). \label{fig:ref-grad-desc}}
\end{minipage}
\hfill
\begin{minipage}{0.38\textwidth}
\includegraphics[width=\textwidth]{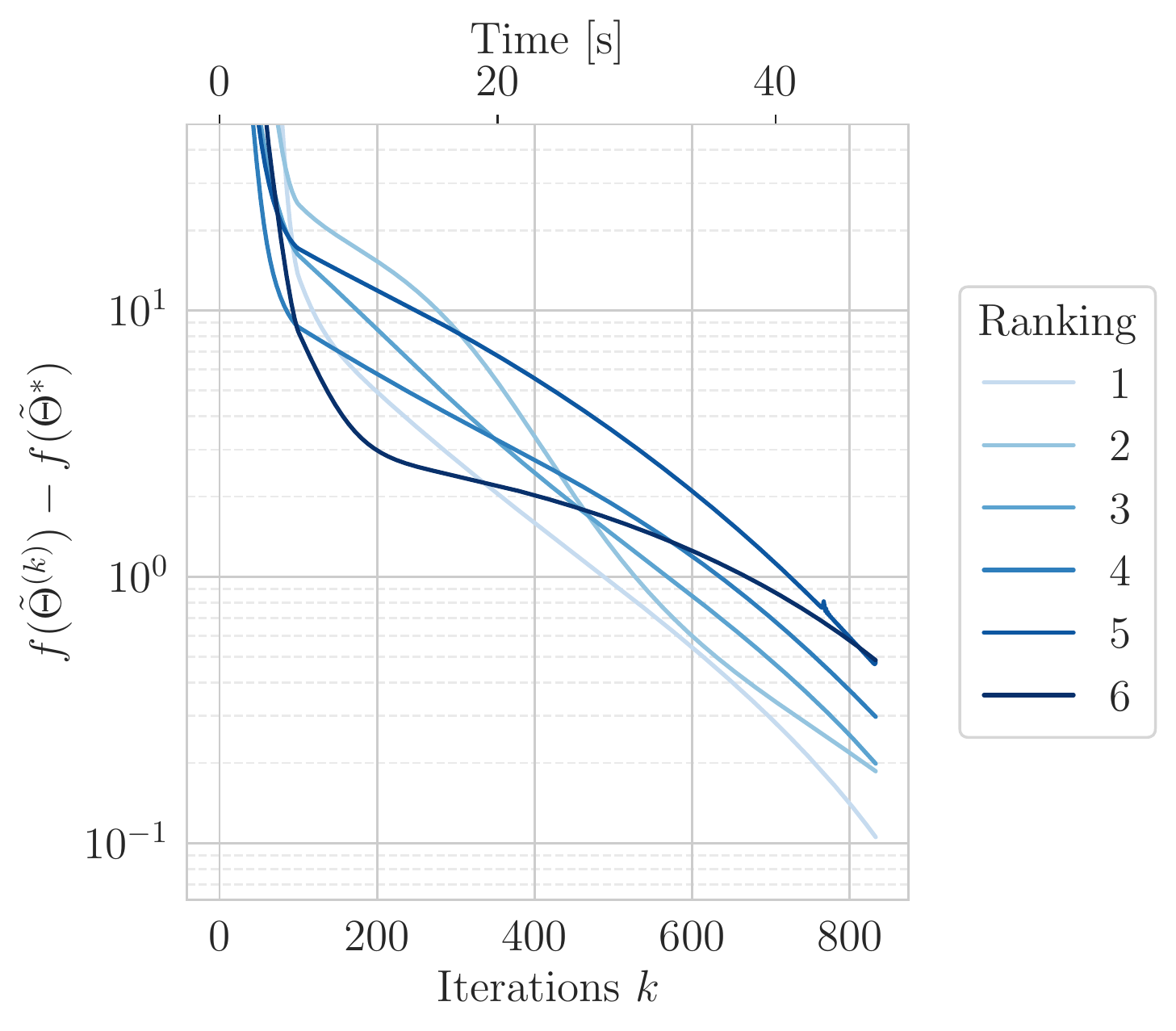}
\caption{Six of the best performing loss trajectories, ranked, over \agu{800} iterations of gradient descent. \agu{We have run 800 iterations (after which the \textit{total} $\chi^2$ difference is $\lesssim 0.5$) solely for demonstration purposes. In our pipeline, however, we terminate the gradient descent at the 300th iteration, because at that stage, the best performing trajectory always ends up at the global minimum.}
Also, with the $y$-axis in $\log$ scale, note the approximate geometric convergence of each solution starting around the 200th iteration, 
after the initial steep descent.
\label{fig:ref-loss}}
\end{minipage}

\end{figure}

There are different approaches to global optimization for non-convex, multimodal functions such as \cref{eqn:obj-fun}.
For example, \texttt{lenstronomy} uses particle swarm optimization (PSO). Although this method is attractive because it requires only the evaluation of the objective function $f$ (and not its gradient), 
a major weakness is that there are ``little to no guarantees'' \citep{sengupta2018} for finding the global, or even local, minimum.
We take the approach of gradient descent, 
which has a number of advantages over 
heuristic optimization techniques such as PSO. 
Most importantly, gradient descent can at least guarantee convergence towards a local minimum. 
Furthermore, when close enough to a local minimum, $\tilde{\Theta}^*$,
gradient descent approximately achieves a geometric convergence rate. 
That is, $f(\tilde{\Theta}^{(k)}) - f(\tilde{\Theta}^*)$ is upper bounded by $\order{\beta^k}$, where $\beta < 1$ and $\tilde{\Theta}^{(k)}$ is the candidate solution on the $k$th iteration, called the $k$th iterate \citep{nesterov2014}.
To achieve the global optimum, we disperse a large number of samples $n_{MAP}$ throughout a wide region of the parameter space and carry out gradient descent on each of these samples \citep{marti2003,gyorgy2011}. 
Since the iterates of these samples will quickly and reliably converge towards local minima, we only need to ensure that $n_{MAP}$ is large enough such that at least one of the samples 
reaches sufficient proximity of the global minimum. \agu{The loss function $f(\tilde{\Theta})$ is multimodal in the sense that there exist multiple local minima\footnote{\agu{However, the posterior is \textit{not} multimodal in the sense that each of these local minima has vanishing posterior density relative to the global mode. Although each of these local minima have a large effect on the performance of MAP, for sampling, they are irrelevant.}}, but} we find that a moderate number of samples, $n_{MAP} = 300$ and $K_{MAP}=300$ iterations of gradient descent is \agu{sufficient} for consistent identification of the global optimum. \agu{After $K_{MAP}$ iterations, we take the best of the $n_{MAP}$ samples to be the MAP estimate $\tilde{\Theta}^*_{MAP}$.}
As we show in \cref{fig:ref-grad-desc}, many samples, even some that start far away from the global minimum, converge to the neighbourhood of the correct solution. If the gradient descent were run for more iterations, each of them would reach the same, globally optimal solution \agu{(see \cref{fig:ref-loss})}. \agu{Typically $\sim 5\%$ (and at least $1\%$, for systems with a low signal-to-noise ratio) of the samples converge to the global optimum, with the other samples eventually converging to local optima.} This demonstrates the robustness of the multi-starts gradient descent method. 

Each of the $n_{MAP}$ samples are initialized by sampling from the prior: $\tilde{\Theta}_i^{(1)} \sim \tilde{p}(\tilde{\Theta})$ for $i=~1,\ldots, n_{MAP}$. 
Finally, similar to the training of neural networks, we use the Adam optimizer \citep{kingma2017} with an initial large learning rate\footnote{The learning rate is used as a multiplier for the gradient when updating parameters in gradient descent, with the simplest implementation being $\tilde{\Theta}^{(k+1)}_i=\tilde{\Theta}^{(k)}_i-\alpha \grad f(\tilde{\Theta}^{(k)}_i)$. The Adam optimizer is slightly more complex, and rescales the components of the gradient before applying the update. We refer readers to \citet{kingma2017} for a more detailed description of the Adam update rule.} $\alpha=10^{-2}$ to accelerate 
``learning'' and escape spurious local minima, and decay it to $\alpha=10^{-3}$ over 300 iterations to help the optimization converge and avoid instabilities \citep{you2019}. This is reflected in the loss trajectories shown in \cref{fig:ref-loss}, which show a period of rapid improvement in the first \agu{$\sim 200$} iterations toward the neighborhood that surrounds the global mode, followed by approximate geometric convergence.

As noted in \S\ref{sec:ad}, with AD,
the gradient can be calculated at virtually no additional computational cost. 
Furthermore, after initializing $n_{MAP}$ samples, the optimization can be done simultaneously by virtue of our code's parallelization. 
Combining these two performance enhancements, this step of our modeling pipeline is fast: 
to find the global optimum, it takes just \ag{17 seconds (see \S\ref{sec:pipeline-sum}, \cref{tab:pipeline})} to run 300 iterations of gradient descent with $n_{MAP}=300$ (\cref{fig:ref-grad-desc}). 

\clearpage
\subsection{Variational Inference}\label{sec:vi}
After finding the MAP estimate, $\tilde{\Theta}^*_{MAP}$, it is \agu{necessary} to do an intermediate step of analysis before sampling via Monte Carlo. 
Specifically, we estimate the posterior covariance matrix of the lens parameters $\Sigma$. \aguu{This covariance estimate plays an important auxiliary role for Monte Carlo sampling: it sets a scale for each of the parameters that is used to define a proposal distribution in the sampling step, as we will show in \S\ref{sec:hmc}.}
We find this covariance estimate by using variational inference \citep[VI;][]{blei2017} to fit a multivariate normal $\mathcal{N}(\tilde{\mu},\tilde{\Sigma})$ (called the ``surrogate'' posterior) with a probability density $\tilde{q}(\tilde{\Theta}; \tilde{\mu}, \tilde{\Sigma})$ to the true posterior.\footnote{\agu{This roughly corresponds to calculating the posterior mode and evaluating the Hessian of the log posterior density around the mode.}} This means minimizing the Kullback-Leibler (\kl) divergence between the two distributions:
\begin{equation}
    \tilde{\mu}^*_{VI}, \tilde{\Sigma}^*_{VI} = \underset{\tilde{\mu}, \tilde{\Sigma}}{\text{argmin}} \ \kl (\tilde{q}(\tilde{\Theta}; \tilde{\mu}, \tilde{\Sigma}) \mid \mid \tilde{p}(\tilde{\Theta} \mid \mathcal{I}_{obs})) \label{eqn:vi}
\end{equation}
Since the posterior density $\tilde{p}(\tilde{\Theta} \mid \mathcal{I}_{obs})$ is intractable, we decompose the \kl:
\begin{equation}
\begin{split}
    \kl (\tilde{q}(\tilde{\Theta}; \tilde{\mu}, \tilde{\Sigma}) \mid \mid \tilde{p}(\tilde{\Theta} \mid \mathcal{I}_{obs})) &= \mathbb{E}_{\tilde{\Theta} \sim \mathcal{N}(\tilde{\mu}, \tilde{\Sigma})}\qty[\log \tilde{q}(\tilde{\Theta}; \tilde{\mu}, \tilde{\Sigma})-\log \frac{\tilde{p}(\mathcal{I}_{obs}, \Theta)}{p(\mathcal{I}_{obs})}] \\ &= \underbrace{\mathbb{E}_{\tilde{\Theta}}[\log \tilde{q}(\tilde{\Theta}; \tilde{\mu}, \tilde{\Sigma})-\log \tilde{p}(\mathcal{I}_{obs},\tilde{\Theta})]}_{\text{ELBO loss}} + \underbrace{\mathbb{E}_{\tilde{\Theta}}[\log p(\mathcal{I}_{obs})]}_{\text{Independent of $\tilde{\mu},\tilde{\Sigma}$}} \label{eqn:elbo}
\end{split}
\end{equation}
where $\mathbb{E}_{\tilde{\Theta}}$ denotes the expectation value with respect to the surrogate posterior $\tilde{q}$. 
Therefore, minimizing the \kl divergence is equivalent to minimizing the evidence lower bound (ELBO). 
This is tractable since $\tilde{p}(\mathcal{I}_{obs},\tilde{\Theta})$ \agu{can be expressed as the product of the prior $\tilde{p}(\tilde{\Theta})$ and likelihood $\tilde{p}(\mathcal{I}_{obs} \mid \tilde{\Theta})$, each of which are readily available}. The gradient of this loss is also expressible as an expectation \citep{ranganath2013}:
\begin{equation}
    \grad_{\tilde{\mu}, \tilde{\Sigma}} \kl = \grad_{\tilde{\mu}, \tilde{\Sigma}} \text{ELBO} = \mathbb{E}_{\tilde{\Theta}}\qty[\qty(\log \tilde{q}(\tilde{\Theta}; \tilde{\mu}, \tilde{\Sigma})-\log \tilde{p}(\mathcal{I}_{obs}, \tilde{\Theta})) \grad_{\tilde{\mu},\tilde{\Sigma}} \log \tilde{q}(\tilde{\Theta}; \tilde{\mu}, \tilde{\Sigma})] \label{eqn:elbo-grad}
\end{equation}
In practice, at each iteration, we use AD to calculate $\grad_{\tilde{\mu},\tilde{\Sigma}} \log \tilde{q}(\tilde{\Theta}; \tilde{\mu}, \tilde{\Sigma})$ and approximate the expectation in \cref{eqn:elbo-grad} with a finite number of samples\footnote{This way of doing variational inference is sometimes termed stochastic variational inference \citep{hoffman2013} because of the stochasticity induced by the finite number of samples $n_{VI}$ used at each iteration.} $n_{VI}$ drawn from $\mathcal{N}(\tilde{\mu}, \tilde{\Sigma})$. This forms an estimator for the true gradient $\grad_{\tilde{\mu},\tilde{\Sigma}} \kl$, which we use to do gradient descent with the Adam optimizer to minimize the ELBO in \cref{eqn:elbo}.
We note that the covariance matrix $\tilde{\Sigma}$ is constrained to be positive semi-definite. In keeping with our method of using unconstraining bijectors in \S\ref{sec:ad}, we use a Cholesky bijection mapping unconstrained real vectors to positive semi-definite matrices, and optimize over this unconstrained space.\footnote{The bijector uses the fact that any covariance matrix can be written in terms of its Cholesky decomposition $\Sigma=L L^T$, where the Cholesky factor $L$ is a lower triangular matrix with a non-negative diagonal. The unconstrained space of real vectors $\mathbb{R}^{d(d+1)/2}$ is then mapped to a covariance matrix by first reshaping the vector into a lower triangular matrix, exponentiating the diagonal entries (which defines a valid Cholesky factor $L$), then multiplying $LL^T$ to find the corresponding covariance matrix.} We initialize our VI with the MAP estimate $\tilde{\mu}^{(1)}=\tilde{\Theta}^*_{MAP}$ and a diagonal covariance matrix $\tilde{\Sigma}^{(1)}=10^{-6} \mathbb{I}$\agu{, based on the intuition that it will be easier for VI to approximate the true covariance starting from an underestimate.}
\agu{Based on a coarse optimization for the reference system,} we run VI for $K_{VI}=1000$ iterations using $n_{VI}=500$, with the learning rate being increased quadratically from $\alpha=0$ to $\alpha=10^{-3}$ over 500 iterations.\footnote{We start from $\alpha=0$ to allow Adam to adjust its first and second order moment estimates \ag{\citep[see][]{kingma2017}}.} This slow increase in learning rate is to avoid initial instabilities in the optimization that may result from the crude initial guess for the covariance $\tilde{\Sigma}^{(1)}$. In our TensorFlow implementation of the modeling pipeline, we use the TensorFlow Probability methods for variational inference, whereas in our JAX implementation, we calculate \cref{eqn:elbo-grad} directly.

The resulting best-fit distribution is only an approximation, since the true posterior is not necessarily Gaussian. \cref{fig:archtypes} show an instance of a posterior with non-Gaussian marginals, indicating the true posterior is not Gaussian. Even when the marginals of the posterior appear to be Gaussian, this does not necessarily imply the full posterior is jointly Gaussian (see \cref{fig:ref-corner} and \citet{dutta2014b}). This is consistent with the fact that the VI posterior does not always exactly agree with the true posterior (i.e., HMC samples). Nonetheless, we find that the VI covariance matrix is almost always a sufficiently good estimate of the true covariance. 


\clearpage
\subsection{Hamiltonian Monte Carlo}\label{sec:hmc}
For the last step, sampling, we will use Hamiltonian Monte Carlo \citep[HMC;][]{duan1987a, neal2012a}. 
HMC relies on gradient information about the posterior distribution, and is known to have several advantages over gradient-free MCMC samplers (e.g., \emcee). 
Conveniently, this gradient information is readily available via AD. 
Using HMC, we achieve highly efficient sampling.
These results are shown \S\ref{sec:hmc-result}. 
Furthermore, we compare the performance of HMC and the widely used \emcee sampler in \S\ref{sec:hmcvemcee} and show that in high-dimensional spaces HMC is strongly preferred.

\subsubsection{Sampler Configuration}\label{sec:hmc-result}
We sample from the posterior $\tilde{p}(\tilde{\Theta} \mid \mathcal{I}_{obs})$ in the unconstrained parameter space with HMC and convert back to samples of the \ag{physical} parameters $p(\Theta \mid \mathcal{I}_{obs})$ using the bijector $g$. 
We initialize $n_{HMC}$ chains\aguu{\footnote{\aguu{We use multiple chains, $n_{HMC}=50$, first because our framework naturally lends itself to parallelization, so it is more efficient to sample for $750$ iterations using $50$ chains rather than sampling for $750 \cdot 50$ iterations using one chain. Using multiple chains is also necessary to evaluate sampling diagnostics such as $\hat{R}$ (see end of this section).}}} by sampling from the surrogate posterior calculated by VI (namely, $\mathcal{N}(\tilde{\mu}^*_{VI},\tilde{\Sigma}^*_{VI})$), 
and as is typical for Monte Carlo samplers, run $n_{burn}$ burn-in steps before sampling $n_{sample}$ times.

Correctly configuring the HMC sampler is important to realizing its advantages. 
There are three hyperparameters in HMC: the step size $\epsilon$, number of leapfrog steps $L$, and the mass matrix $\tilde{M}$ that defines the momentum distribution. 
Much work has been done to adaptively set the first two hyperparameters $\epsilon$ and $L$. We use the methods introduced by \citet{hoffman2014,hoffman2021}, as implemented by TensorFlow Probability, to automatically tune (``autotune'') the hyperparameters during the first 80\%\footnote{As recommended by Tensorflow Probability (see \url{https://www.tensorflow.org/probability/api_docs/python/tfp/mcmc/DualAveragingStepSizeAdaptation}).} of burn-in phase. They are then fixed \agu{in the remaining} burn-in steps, since their adaptation generally prevents chains from reaching the stationary distribution (hence the use of only \ag{the first} 80\% of burn-in steps for adaptation). Part of this tuning is adjusting the step size $\epsilon$ to achieve a target acceptance probability for each proposal: a step size that is too small will result in slower sampling, but a step size that is too large will result in too many rejected proposals. Optimal values for the target acceptance probability range between 0.6 to 0.8 \citep{betancourt2018}; we use 0.75. The remaining hyperparameter, the mass matrix $\tilde{M}$, \agu{defines the momentum distribution, and this provides a way to inform} the HMC algorithm about the scales and correlations of the parameters. We can significantly improve the sampling efficiency by setting $\tilde{M}$ to be the inverse covariance matrix of the posterior \citep[][Chap. 5]{brooks2011} --- this is called ``preconditioned'' HMC. 
This was the main purpose of the previous section, \S\ref{sec:vi}: 
we set $\tilde{M}=(\tilde{\Sigma}^*_{VI})^{-1}$, the inverse of the inferred covariance matrix from VI
as defined by \cref{eqn:vi}. \agu{There have been proposals \citep[e.g.,][]{sountsov2021} to adapt the mass matrix $\tilde{M}$ on the fly during the burn-in steps, which may render the VI step irrelevant. For now, these methods are not yet well-tested (however, they may be incorporated in future work), hence VI remains a necessary step of our modeling pipeline. Note that a secondary use of the VI step is that each of the $n_{HMC}$ chains are initialized (see \cref{tab:pipeline}) by sampling from the VI posterior, $\mathcal{N}(\tilde{\mu}_{VI}^*, \tilde{\Sigma}_{VI}^*)$.} 

We show in \cref{fig:demo-posterior-samples} the posterior samples for our reference system generated using $n_{HMC}=50$ chains and $n_{burn},n_{sample}=250,750$. \agu{As with the VI step, these hyperparameters were roughly tuned on the reference system.}
We report two metrics that are widely used in the statistics literature to measure the degree to which our sampler has converged. These are known as the effective sample size (\ess) and potential scale reduction factor (PSRF), $\hat{R}$ \citep{gelman1992}. The former measures the effective number of independent samples we have drawn from the posterior by accounting for autocorrelation within each chain, and the latter is the ratio of the average within-chain variance to the variance of the pooled samples across all chains. A large \ess and a $\hat{R}$ that is close to 1 indicates convergence has been achieved (in \citet{gelman1992}, it is suggested that an appropriate condition is $\hat{R} < 1.1$).

\begin{minipage}{\linewidth}
\makebox[\linewidth]{
    \includegraphics[width=0.95\textwidth]{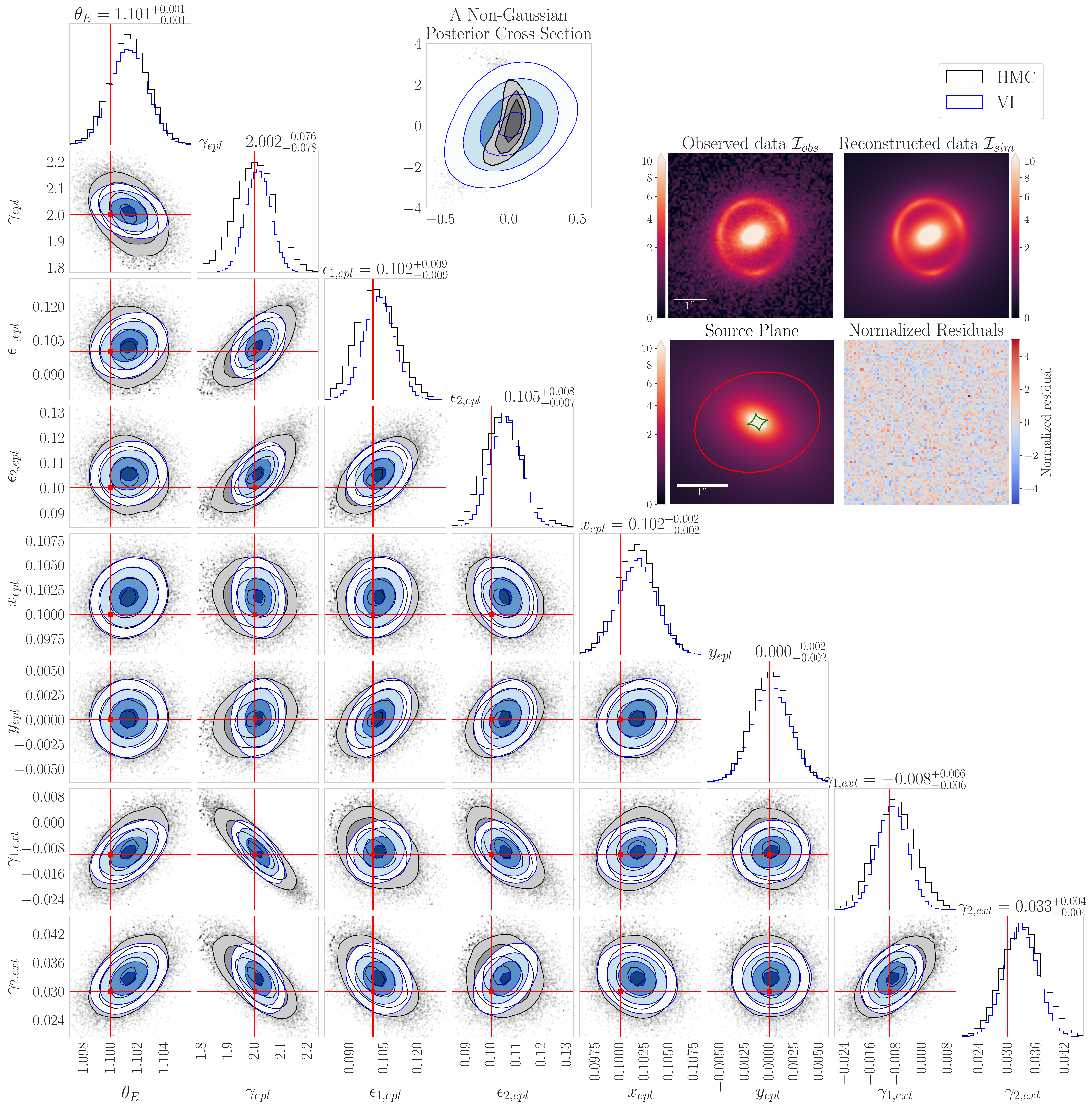}}
    \captionof{figure}{A corner plot of the posterior samples for the reference system (see \cref{eqn:sim-dist} for the definition of each parameter). We show only the lensing parameters, marginalizing out all light parameters \aguu{(marginalized posteriors for all 22 parameters are shown in \cref{fig:demo-posterior-samples})}.
The posterior samples are first obtained in unconstrained space from both VI and HMC, and then converted to physical parameters by applying the bijector $g$ (see text). 
\agu{The samples and $0.5$, $1$, $1.5$, $2$ $\sigma$ contours (corresponding to roughly $12\%$, $39\%$, $68\%$, $86\%$ of the probability mass),} for both VI and HMC are shown in blue and grey, respectively\aguu{, and the ground truth is in red}.
In the top right inset, we show the model reconstructed image, \aguu{residuals (normalized by the square root of the noise variance map)}, and the reconstructed source (together with the caustic shown in green and critical curve in red) using the Bayesian mean estimate. \agu{Despite the approximately Gaussian marginal distributions on the corner plot for the true posterior, we show a random cross section of the posterior that is banana-shaped (top inset), demonstrating that the full posterior is not perfectly Gaussian (see text). Since the VI ansatz is a multivariate Gaussian, this is consistent with the fact that the marginals for the VI posterior do not entirely coincide with those of the true posterior.} \label{fig:ref-corner}
}
\end{minipage}

\begin{minipage}{\linewidth}
\makebox[\linewidth]{
    \includegraphics[width=\textwidth]{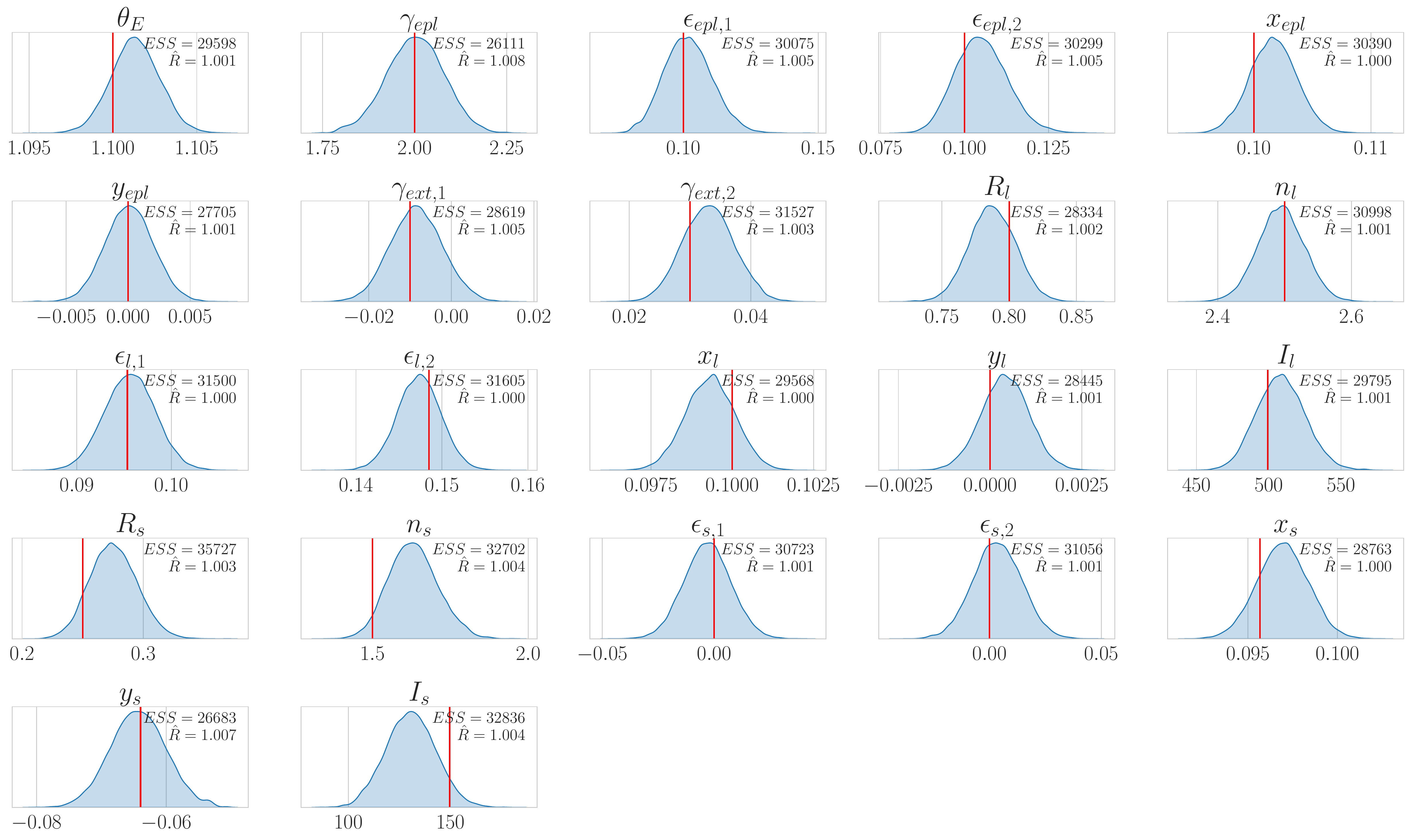}}
    \captionof{figure}{Marginalized posterior samples for all 22 parameters of the reference system. As in \cref{fig:ref-corner}, the \agu{ground truth} (input values) are marked in red. We report the effective sample size (\ess) and potential scale reduction factor $\hat{R}$ for our posterior samples.
    Note that for each parameter, $\ess > 26000$ and $\hat{R} < 1.01$. 
    This is achieved with just under 36~seconds of HMC sampling (\cref{tab:pipeline}).\label{fig:demo-posterior-samples}}
\end{minipage}

\subsubsection{Comparison of HMC and \emcee}\label{sec:hmcvemcee}
\citet{foreman2013a} implemented an affine-invariant ensemble sampler, \emcee.
\xh{It is a popular MCMC algorithm in astrophysics.}
This is the sampler that \texttt{lenstronomy} uses.
Here we compare the performance of HMC with \emcee, 
which is gradient-free, by applying both to the reference system (\cref{fig:ref-sys}). \agu{To make the comparison as fair as possible, for \emcee sampling, we initialize the sampler with the \lenstr recommended configuration, as detailed in \citet{birrer2021}, and for our HMC sampling, we initialize the sampler as detailed in \cref{tab:pipeline}. Furthermore, we run our pipeline on a single A100 GPU and \emcee on a single CPU. \lenstr uses uniform priors for each parameter whereas we use the prior described in \cref{eqn:sim-dist}. However, we have found that the difference in priors has virtually no effect on the sampling results. Finally, for both modeling pipelines, we use the supersampling factor $k_{super}=2$ and the PSF shown in \cref{fig:ref-sys}.}

We take two axes of comparison between HMC and \emcee. 
First, we observe that our sampling process is significantly more efficient than \emcee, as evidenced by the rate at which HMC generates independent samples, \agu{$\sim 40 \ \ess/\text{iter}$ ($\sim 300 \ \ess/\text{sec}$, on a single A100 GPU)} 
whereas for \texttt{emcee} it is \agu{$\sim 0.2 \ \ess/\text{iter}$ ($\sim 0.04 \ \ess/\text{sec}$)} (see \cref{fig:hmc-comp}). Second, we compare the convergence of the two samplers. We find that although both sampling methods agree in terms of their central values, they exhibit dramatically different convergence behavior. 
In \cref{fig:hmc-comp}, we show that individual \texttt{emcee} chains tend to devolve to random walks. \agu{This random walk behavior manifests itself in three ways. First, \emcee} makes slow progress exploring the posterior, whereas HMC draws virtually independent samples each iteration, traversing the posterior very efficiently. \agu{Second, compared with HMC,} we observe high inter-chain variance in \texttt{emcee}, evidenced qualitatively by the differing marginal distributions for each of the individual chains, and quantitatively by the substantially higher $\hat{R}$ for \texttt{emcee} (\cref{fig:hmc-comp}).
\agu{Third,} in \cref{fig:autocorr} we \agu{find} that the autocorrelation time for HMC is 
much lower than that of \texttt{emcee}: within just 10 iterations, the autocorrelation shrinks to negligible levels, compared to \texttt{emcee}, which has a characteristic autocorrelation lag of $\sim 300$. The empirical autocorrelation at lag $\tau$ for a single MC chain $\qty{f_n \mid n=1,\ldots,N}$ is defined by \citep{sokal1996}:

\begin{equation}
    \begin{gathered}
    \hat{\rho}(\tau)=\hat{c}(\tau)/\hat{c}(0) \qc{\text{where}} \\
    \hat{c}(\tau)=\frac{1}{N-\tau} \sum_{n=1}^{N-\tau} (f_n-\mu_f)(f_{n+\tau}-\mu_f) \qq{and} \mu_f=\frac{1}{N} \sum_{n=1}^N f_n.
    \end{gathered}
\end{equation}
We emphasize that this autocorrelation is independent of the iteration number. 
That is, burn-in does not remove autocorrelation, nor does running a chain for a very long time.

\begin{minipage}{\linewidth}
\makebox[\linewidth]{
    \includegraphics[width=0.7\textwidth]{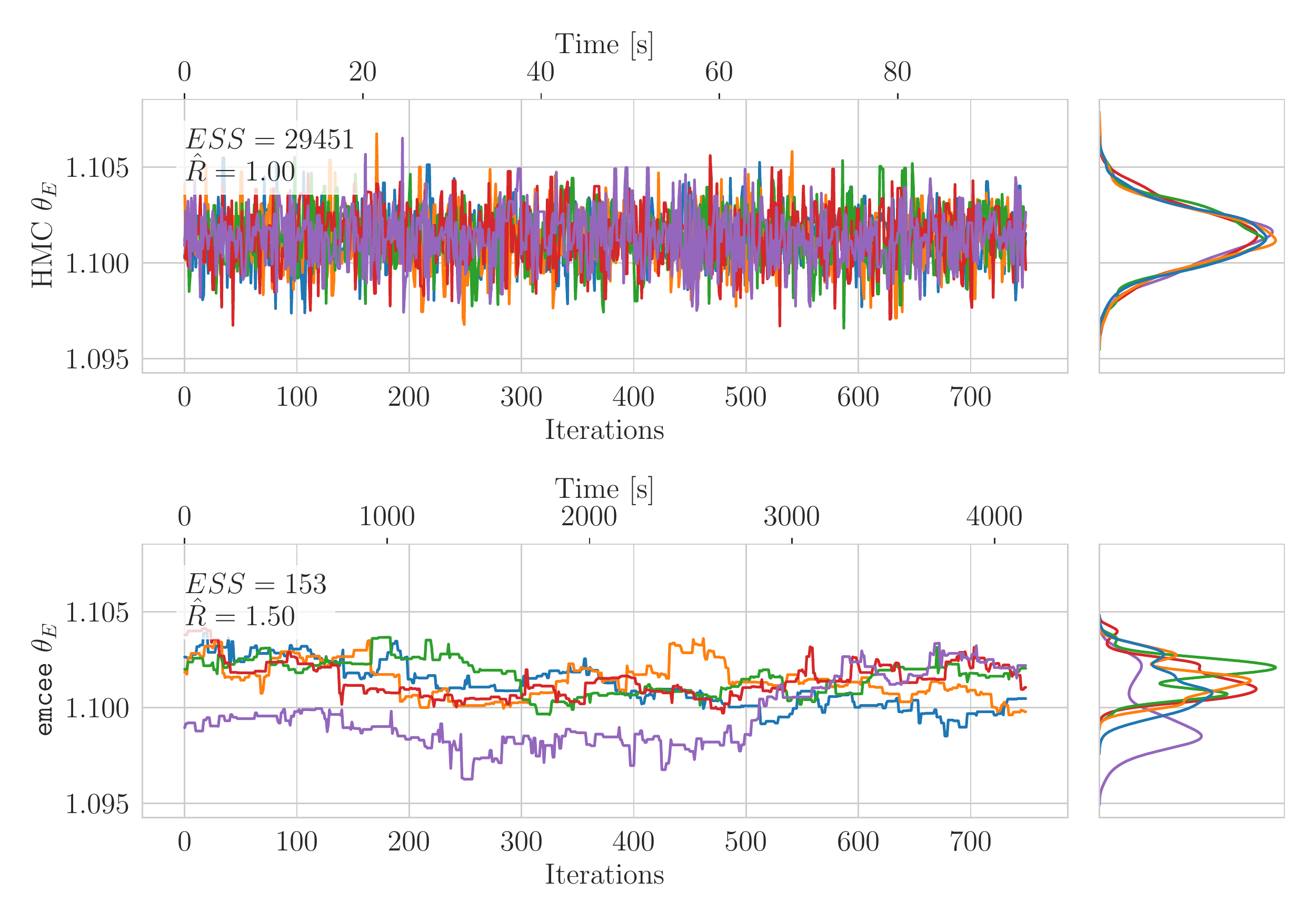}}
    \captionof{figure}{A comparison of HMC with \texttt{emcee} performance for the Einstein radius. 
    We illustrate 5 randomly chosen chains (from 50 total chains) for both HMC and \texttt{emcee} over 750 sampling iterations. The times
    shown are for HMC run on a \textit{single} A100 GPU and \emcee run on a modern CPU. Note the poor interchain mixing in \emcee, which leads to a high $\hat{R}$.  In particular, for \emcee, $\hat{R}$ is much higher than the recommended value of 1.1.
    This is representative behavior for all other physical parameters.}
    \label{fig:hmc-comp}
\end{minipage}

\begin{minipage}{\linewidth}
\makebox[\linewidth]{    \includegraphics[width=0.75\textwidth]{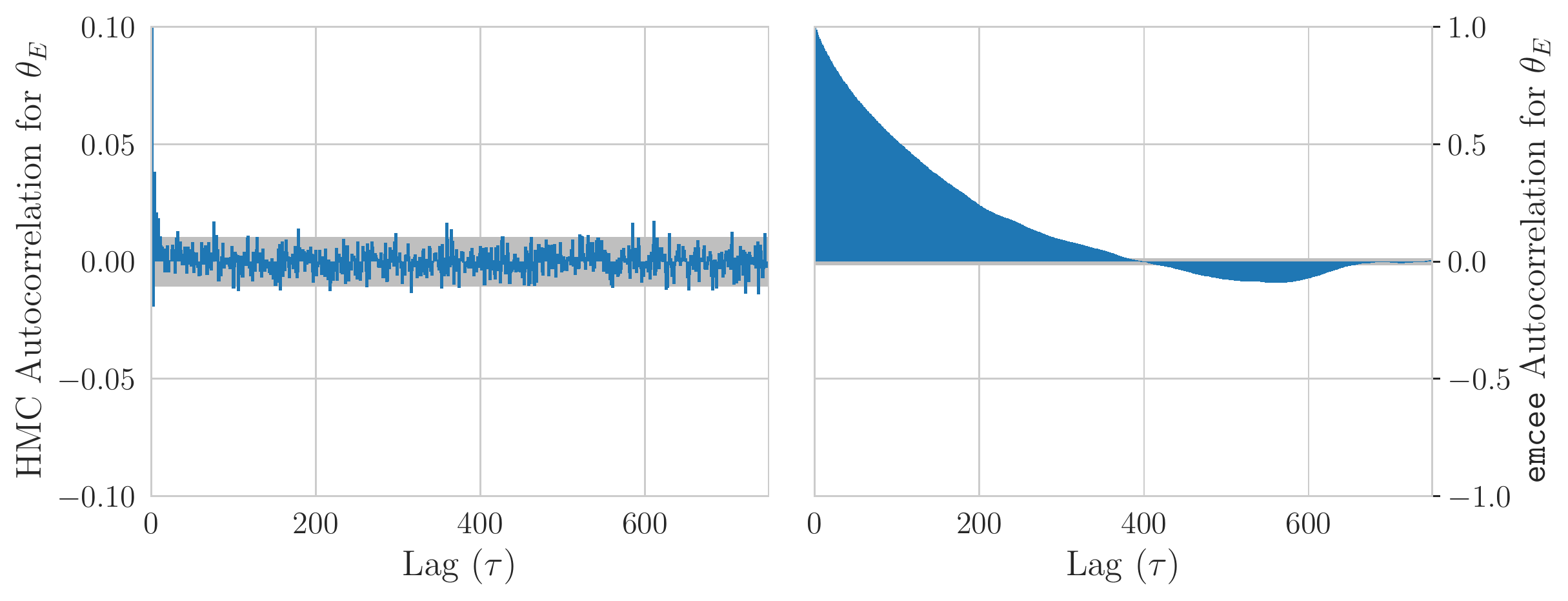}}
    \captionof{figure}{Autocorrelation of the Einstein radius, $\hat{\rho}(\tau)$, using HMC and \texttt{emcee} (see text). 
    This is representative behavior for all other physical parameters. 
    Note the vertical scale for HMC is one tenth of that of \texttt{emcee}.}
    \label{fig:autocorr}
\end{minipage}

\ag{Our investigation of \emcee revealed undesirable characteristics even for low to moderate dimensional spaces, as shown above in the case of the reference system with $22$ parameters. While in this regime, it is possible that through tuning and longer sampling time, higher quality convergence can still be achieved using \emcee, \citet{betancourt2018} pointed out that in higher dimensions, any gradient-free sampler is likely to be much less efficient compared with HMC.} Furthermore, \cite{huijser2017} found that in moderate ($\sim 50$) to high ($> 100$) dimensions, affine-invariant ensemble samplers (such as \emcee) can have more severe problems. \agu{They showed that for high dimensional posteriors, in addition to slow convergence, an \aguu{affine invariant sampler} can misleadingly appear to converge even when it has not.}
In strong lens modeling, it is critical to avoid \agu{this pernicious behavior}, since models for high resolution observed data that use complex light profiles such as shapelets \citep{birrer2015}, wavelets \citep{galan2021a}, or pixelization \citep{nightingale2021} can easily have $\gtrsim 50$ parameters. 
\xh{The modeling of perturbations to the smooth lensing potential, 
whether due to dark matter subhalos or line-of-sight halos will require even more.} 
If we wish to fit these sophisticated models to observed data, we must be able to do robust inference in spaces of moderate to high dimensions. 

\subsection{Pipeline Summary and Hyperparameter Settings}\label{sec:pipeline-sum}
Our pipeline is a sequence of three steps
with the ultimate goal of producing a collection of samples from the posterior distribution from which robust statistical inferences can be made. 
We summarize these three steps in \cref{tab:pipeline}, and report the hyperparameter and initialization settings that we used for the reference system.

From our experience of using \texttt{lenstronomy}, the PSO initialization usually needs to be at least somewhat close to the optimum. With multi-start gradient descent, we \aguu{find this to be unnecessary}. While samples that start near the optimum are virtually assured to reach it, as expected, those that start far away can often succeed as well (see \cref{fig:ref-grad-desc}). This suggests that multi-start gradient descent has a much weaker dependence on initialization than PSO. In the next section, we will show the application of our pipeline to 100 simulated systems. We find that the MAP initialization in \cref{tab:pipeline} does not need to be adjusted to successfully model these systems, providing further evidence that multi-start gradient descent is not sensitive to initialization\agu{, so long as the prior is broad and $n_{MAP}$ is sufficiently large}. The remaining initializations (for VI and HMC) do not need to be changed either. Furthermore, the hyperparameters in \cref{tab:pipeline} that were chosen for modeling the reference system have also been found to suffice for the 100 simulated systems in \S\ref{sec:results}.

\begin{deluxetable}{l m{3.1cm} r l m{2.8cm} c h}
\tabletypesize{\footnotesize}
\tablecaption{Summary of the modeling pipeline\label{tab:pipeline}}
\tablehead{
\colhead{\textbf{Step}} & \colhead{\textbf{Output}} & \multicolumn2c{\textbf{Hyperparameters}} & \colhead{\textbf{Initialization}} & \colhead{\textbf{Execution time}}}
\startdata
\arrayrulecolor{gray}
& \multirow{3}{=}{An estimate of the posterior mode $\tilde{\Theta}^*_{MAP}$.} & $K_{MAP}$: & 300 & \multirow{3}{=}{\centering $\tilde{\Theta}_i^{(1)} \sim \tilde{p}(\tilde{\Theta})$ \\ $i=1,\ldots,n_{MAP}$} & & \multirow{3}{*}{\centering 50 min.} \\
1. MAP (\S\ref{sec:map}) & &  $n_{MAP}$: & 300 & & 17 sec. \\
& &  $\alpha$: & $10^{-2} \overset{\text{lin,300}}{\longrightarrow} 10^{-3}$ & \\
\hline
& \multirow{3}{=}{An estimate of the posterior mean $\tilde{\mu}_{VI}^*$ and covariance $\tilde{\Sigma}_{VI}^*$.} & $K_{VI}$: & 1000 & \multirow{3}{=}{\centering $\tilde{\mu}^{(1)}=\tilde{\Theta}^*_{MAP}$ \\ $\tilde{\Sigma}^{(1)}=10^{-6} \mathbb{I}$} & & \multirow{3}{*}{\centering N/A} \\
2. VI (\S\ref{sec:vi}) & & $n_{VI}$: & 500 & & 52 sec. \\
& & $\alpha$: & $0 \overset{\text{quad,500}}{\longrightarrow} 10^{-3}$ \\
\hline
& \multirow{5}{=}{Samples drawn from the posterior $p(\Theta \mid \mathcal{I}_{obs})$} &  $n_{burn}$: & $250$ & \multirow{5}{=}{\centering Initialize $n_{HMC}$ walkers by sampling from the VI posterior} & & \multirow{5}{*}{\centering 70 min.} \\
& &  $n_{sample}$: & 750 & \\ 
3. HMC (\S\ref{sec:hmc}) & & $n_{HMC}$: & 50 & & 36 sec. \\
& &  $\epsilon$: & 0.3 & \\
& &  $L$: & 5 & \\
& &  $\tilde{M}$: & $(\tilde{\Sigma}^*_{VI})^{-1}$ & \\
\hline\arrayrulecolor{black}
\textbf{Total} & & & & & \textbf{105 sec.} & \textbf{120 min.}
\enddata

\tablecomments{The hyperparameters for each step are defined in their respective subsections (first column). 
We adopt the notation $\alpha_1 \overset{s,k}{\longrightarrow} \alpha_2$ to indicate a learning rate that changes from $\alpha_1$ to $\alpha_2$ over $k$ iterations with a polynomial schedule $s$ (in our case, linear or quadratic). The rightmost columns indicate typical execution times for each modeling step on four A100 GPUs. \agu{On a single A100 GPU, the runtime is approximately 3.5 times longer (see \cref{fig:hmc-comp}), totaling $\sim 6$ minutes.}}
\vspace{-2em}
\end{deluxetable}

The total execution time \ag{for our reference system} is 6 minutes on a cutting-edge A100 GPU (available through NERSC Perlmutter early access\footnote{\url{https://www.nersc.gov/systems/perlmutter/}}). On a GPU node on Perlmutter, which has 4 A100 GPUs\footnote{Currently, only the JAX implementation of our pipeline supports distributed computing over multiple GPUs, due to the lack of support for distributed computing on TensorFlow (outside of neural networks).}, it takes 105 sec.  (\cref{tab:pipeline}).

\section{Results}
\label{sec:results}
To demonstrate the performance of our lens modeling pipeline,
we simulate a sample of \agu{100} systems using \texttt{lenstronomy} (see \cref{fig:lens-sample}). 
The parameters for these systems are sampled from the simulation distribution defined in \cref{eqn:sim-dist}. \agu{Our prior, also defined in \cref{eqn:sim-dist}, has the same center as the simulation distribution, but has been broadened considerably so that it is less informative.}


\begin{minipage}{0.96\linewidth}
\vspace{1mm}
\makebox[\linewidth]{
\includegraphics[width=\textwidth]{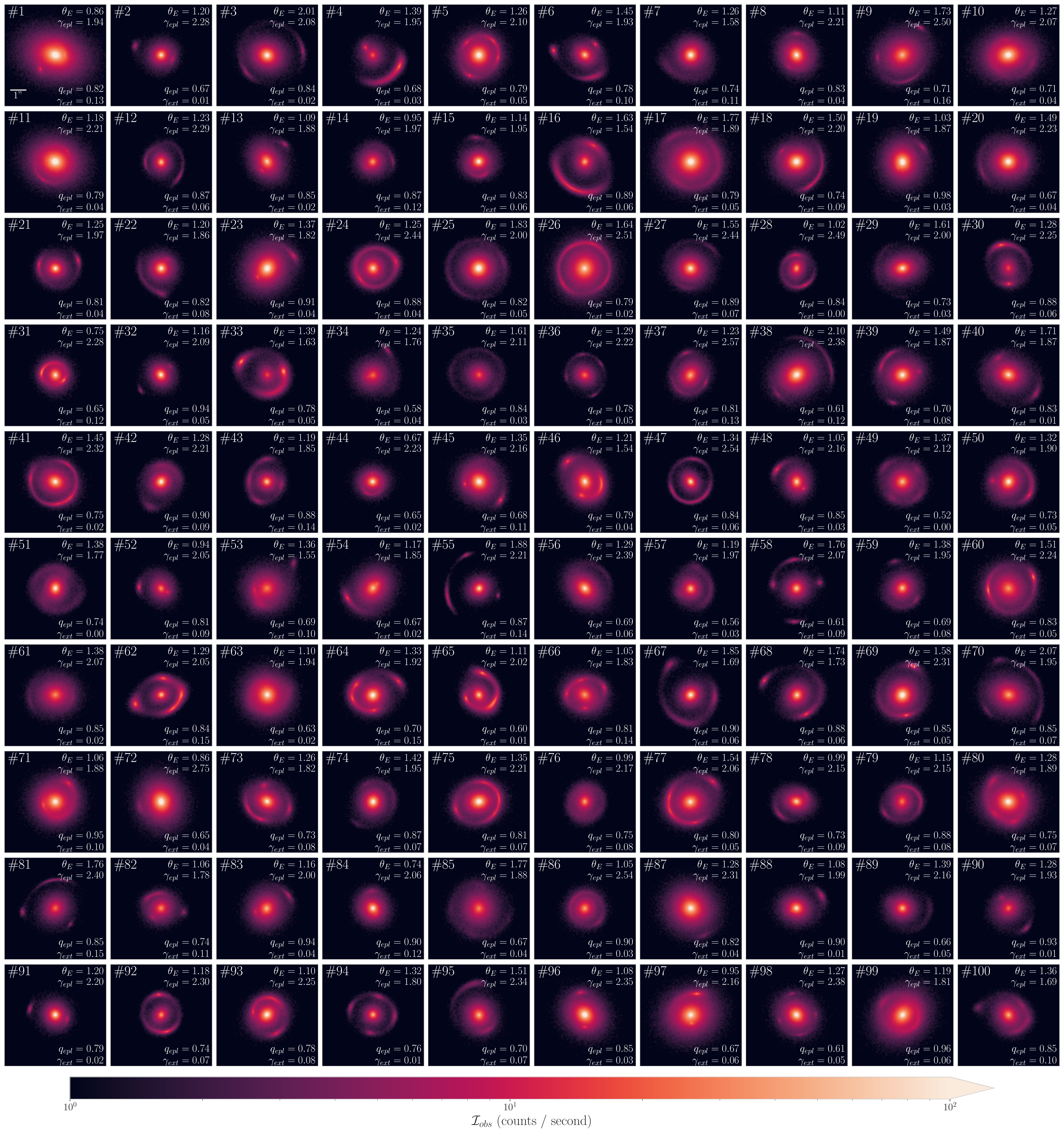}}
    \captionof{figure}
    {
    A sample of \agu{100} lenses simulated using \texttt{lenstronomy}. We include the effects of Gaussian noise with standard deviation $\sigma_{bkg}=0.2$, Poisson shot noise with an exposure time $t_{exp}=100 \ \text{sec}$ with $\mathcal{G}=1$ \aguu{(assuming \hst observations)}, and the PSF (see \cref{fig:ref-sys}). \agu{The pixel scale is $0.065\twopr$ and the cutout size is $5.2\twopr \times 5.2 \twopr$ (80 pixels by 80 pixels).}}
    \label{fig:lens-sample}    
\vspace{2mm}
\end{minipage}

We apply our modeling pipeline as described in \cref{tab:pipeline} to each of these systems and show the excellent agreement with \agu{the ground truth} (input values) in \cref{fig:sim-parity}. \agu{The hyperparameters listed in \cref{tab:pipeline} were roughly tuned (to the appropriate order of magnitude) on the reference system, and left unchanged when modeling the sample of 100 simulated systems.} \ag{The average time to model one simulated system is comparable to the reference system  \aguu{(see \cref{tab:pipeline})}.}
Moreover, we find that our pipeline consistently exhibits favorable MC convergence (\cref{tab:err}): even the largest $\hat{R}$ for any parameter over all \agu{100} simulated systems was $1.017$, and the smallest ESS was $26822$, an order of magnitude higher than the typical value with \emcee.

\begin{minipage}{0.96\linewidth}
\makebox[\linewidth]{
    \includegraphics[width=\textwidth]{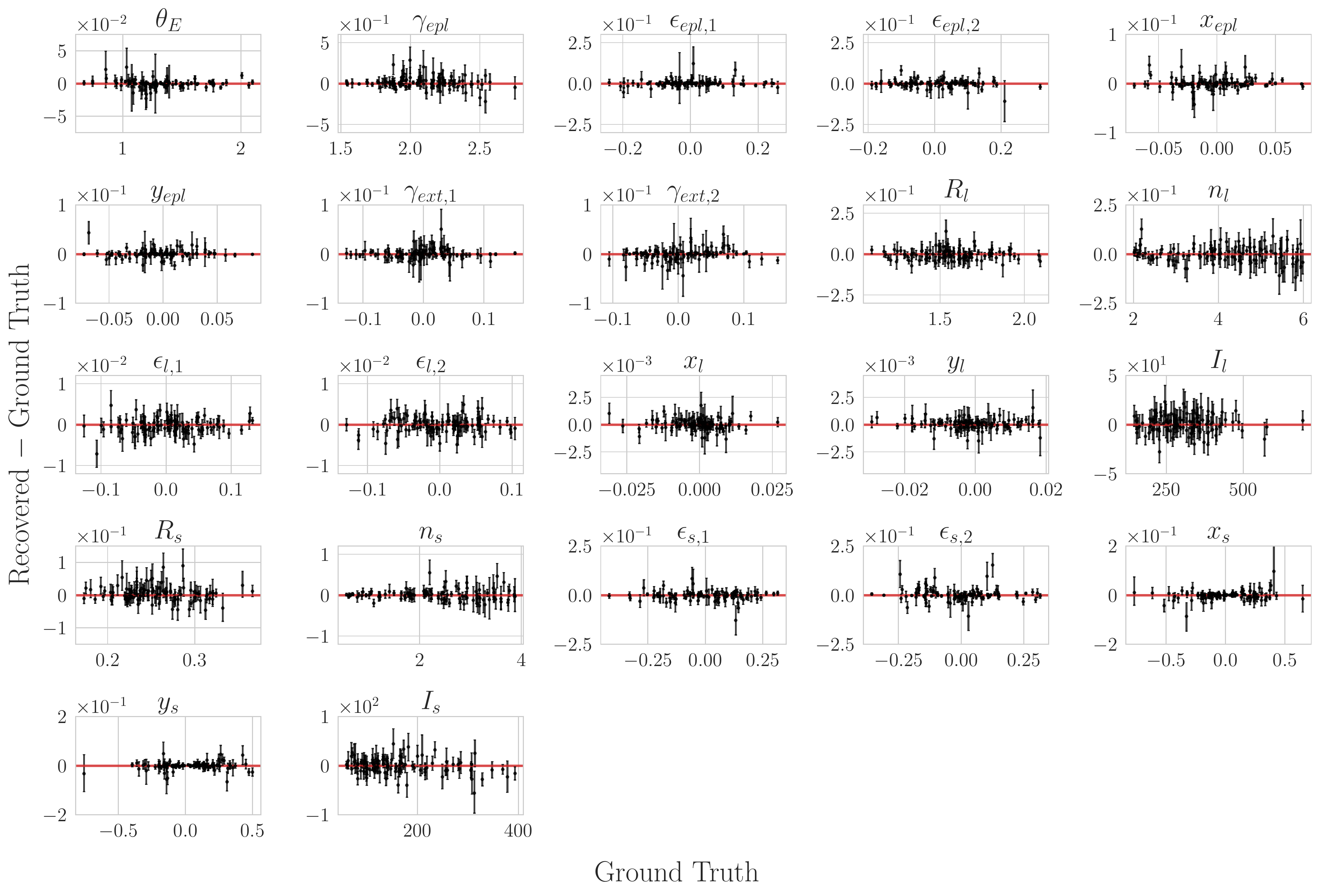}}
    \captionof{figure}{Difference between the \ag{recovered parameters and} \agu{ground truth (input values)} for the \agu{100} simulated systems in \cref{fig:lens-sample}. The points are the mean of the posterior, and the uncertainties correspond to the 68\% highest posterior density interval. \agu{Note that as $n_l$ and $n_s$ increase, their uncertainties increase as well. This is because the light becomes more compact at higher S\'ersic indices, resulting in higher degeneracy between the S\'ersic indices and the half-light radii.}}
    \label{fig:sim-parity}
\end{minipage}

\begin{deluxetable}{lRR|RRRR}[H]
\tablecaption{Summary statistics of lensing parameters\label{tab:err}}
\tablehead{\colhead{Parameter} & \colhead{Mean error} & \colhead{$\mu_z$} & \colhead{$\langle \hat{R} \rangle$} & \colhead{$\max \hat{R}$} & \colhead{$\langle ESS \rangle$} & \colhead{$\min ESS$}}
\startdata
 $\bm{\theta_E}$ &    -0.00026 & -0.04 \pm 0.09 &                      1.001 &            1.013 &                  35465 &        30846 \\
    $\bm{\gamma_{epl}}$ &     0.01608 &  0.12 \pm 0.08 &                      1.001 &            1.017 &                  35407 &        28045 \\
$\bm{\epsilon_{epl,1}}$ &     0.00235 &  0.08 \pm 0.09 &                      1.001 &            1.011 &                  35590 &        29438 \\
$\bm{\epsilon_{epl,2}}$ &    -0.00159 &  0.01 \pm 0.10 &                      1.001 &            1.006 &                  35505 &        31213 \\
         $\bm{x_{epl}}$ &     0.00031 & -0.10 \pm 0.10 &                      1.001 &            1.004 &                  35617 &        33434 \\
         $\bm{y_{epl}}$ &     0.00082 &  0.06 \pm 0.09 &                      1.001 &            1.008 &                  35569 &        32745 \\
  $\bm{\gamma_{ext,1}}$ &     0.00088 &  0.07 \pm 0.09 &                      1.001 &            1.012 &                  35542 &        27926 \\
  $\bm{\gamma_{ext,2}}$ &    -0.00060 & -0.06 \pm 0.09 &                      1.001 &            1.010 &                  35382 &        30456 \\
             $\bm{R_l}$ &    -0.00203 & -0.09 \pm 0.09 &                      1.000 &            1.007 &                  35768 &        33695 \\
             $\bm{n_l}$ &    -0.00321 & -0.05 \pm 0.08 &                      1.000 &            1.008 &                  35777 &        33752 \\
  $\bm{\epsilon_{l,1}}$ &    -0.00038 & -0.21 \pm 0.10 &                      1.000 &            1.003 &                  35444 &        33593 \\
  $\bm{\epsilon_{l,2}}$ &    -0.00006 & -0.01 \pm 0.10 &                      1.001 &            1.003 &                  35495 &        33247 \\
             $\bm{x_l}$ &     0.00003 &  0.03 \pm 0.09 &                      1.001 &            1.003 &                  35687 &        33086 \\
             $\bm{y_l}$ &     0.00006 &  0.13 \pm 0.08 &                      1.000 &            1.003 &                  35641 &        32861 \\
             \aguu{$\bm{I_l}$} &     1.20330 &  0.07 \pm 0.08 &                      1.000 &            1.007 &                  35758 &        33603 \\
             $\bm{R_s}$ &     0.00599 &  0.13 \pm 0.09 &                      1.000 &            1.005 &                  35497 &        32109 \\
             $\bm{n_s}$ &     0.01192 &  0.04 \pm 0.09 &                      1.000 &            1.005 &                  35613 &        32820 \\
  $\bm{\epsilon_{s,1}}$ &    -0.00255 & -0.09 \pm 0.10 &                      1.000 &            1.004 &                  35556 &        31342 \\
  $\bm{\epsilon_{s,2}}$ &     0.00282 &  0.05 \pm 0.10 &                      1.001 &            1.005 &                  35711 &        31810 \\
             $\bm{x_s}$ &    -0.00087 & -0.06 \pm 0.09 &                      1.001 &            1.017 &                  35443 &        26822 \\
             $\bm{y_s}$ &    -0.00092 &  0.03 \pm 0.08 &                      1.001 &            1.017 &                  35470 &        29876 \\
             \aguu{$\bm{I_s}$} &    -0.09109 & -0.05 \pm 0.09 &                      1.000 &            1.003 &                  35580 &        33416
\enddata
\tablecomments{We show the errors for the 22 lensing parameters in \cref{fig:sim-parity}. Mean error denotes the average difference between the recovered parameters and the \agu{ground truth} for the \agu{100} simulated systems.
The notation $z$ denotes errors that have been scaled by the posterior standard deviation. \agu{For example, for a given system, if the posterior mean and variance of the Einstein radius are $\mathbb{E}[\theta_E],\mathbb{V}[\theta_E]$, and the ground truth Einstein radius is $\bar{\theta}_E$, then $z[\theta_E]=(\mathbb{E}[\theta_E]-\bar{\theta}_E)/\sqrt{\mathbb{V}[\theta_E]}$. We report the average (over all 100 systems) scaled error $\mu_z$ for each parameter, and find that they are all consistent with zero bias.} We also report statistics for the MC convergence diagnostics, including the mean and extremal values. Specifically, for any given parameter, the $\max \hat{R}$ and $\min ESS$ values are the largest $\hat{R}$ and smallest $ESS$ for that parameter across all \agu{100} simulated systems.}
\end{deluxetable}

\section{Discussion and Conclusion}\label{sec:conclusion}
\ag{In this work we present a new framework for modeling strong gravitational lenses that is robust, efficient, and scalable to high-dimensional parameter spaces. We achieve this via algorithmic improvements and extensive use of two technologies. For the former, we use multi-start gradient descent in place of PSO, and HMC augmented with VI in place of \emcee. 
For the latter, first,} massive parallel processing on GPUs allows us to simulate thousands of systems at once, orders of magnitude faster than existing lensing codes that use CPUs. 
This fast simulation capability is key 
for efficient forward modeling. 
Second, automatic differentiation \ag{provides access to gradient information that is a highly valuable guide for each step in our pipeline, 
at virtually no \xh{additional computational} cost.} 

\ag{We have demonstrated our pipeline's performance on a large set of simulated systems.} We make a \agu{reasonably general} choice for our lens model (EPL + external shear, with lens and source light modeled with S\'ersic profiles) in this work. 
\ag{But} we emphasize that our modeling methodology is an overarching framework. 
The capabilities described above are applicable to any parameterized lens model. 
For instance, if we opt instead to use shapelets \citep{birrer2015} as a source light model, only $\sim 10\%$ more computation time is needed. 
More importantly, as we \xh{showed} in \S\ref{sec:hmcvemcee}, a gradient-informed modeling pipeline is necessary to do rigorous statistical inference on models with many parameters. 
\agu{Fifty-one of the lensing systems that we discovered in \citet{huang2020a,huang2021a} have been observed with the \textit{Hubble Space Telescope} (ID: 15867; PI: Huang). We will apply the \gigalens framework to model a subset of these systems and report the results in an upcoming publication (Gu et al. in prep.).}

In this work, we have developed the \ag{core} components for a gradient-based lens modeling framework. 
There is much room for expansion within this framework. \agu{For instance, although we did not find significant multimodality in the posterior for the model we consider in this work (i.e., all local modes have vanishing posterior density compared to the global mode), it is unclear whether this will still be the case for more complex lens and source models.} We believe this can be addressed using more advanced samplers (using HMC as a substrate) such as adiabatic Monte Carlo \citep{betancourt2015}, parallel tempering \citep{earl2005}, or annealed importance sampling \citep{neal2001}. The latter is also capable of estimating normalizing constants\ag{, which enables the computation of Bayes factors. This is necessary for model comparison, which is particularly useful for tasks such as the modeling \agu{of subhalos and line-of-sight low-mass halos.}}



\ag{Finally, the execution time can very likely be significantly shortened from the 105~seconds reported in this work via a combination of technological and algorithmic improvements.}
\agu{For the former, we plan to use 8 A100 GPUs, and expect that this will bring the execution time to roughly cut in half, bringing the total time to below 1 minute.} In addition, further improvement on GPU speed is almost a certainty. For the latter, on one hand, advances in mass matrix adaptation for HMC \citep{stan} may allow the VI step to be eliminated, potentially offering up to a factor of two speed \xh{gain}. 
On the other hand, the VI step can be improved to fit the posterior exactly \citep{kingma2017a,papamakarios2018}, allowing HMC to be eliminated from the pipeline. 
This framework and its further improvements make it possible, for the first time, that the $\mathcal{O}(10^5)$ strong lenses expected to be discovered in the next generation surveys can be modeled on a reasonable time scale.

\section{Acknowledgement}\label{sec:acknowledgement}
\aguu{This work was supported in part by the Director, Office of Science, Office of High Energy Physics of the US Department of Energy under contract No. DE-AC025CH11231. This research used resources of the National Energy Research Scientific Computing Center (NERSC), a U.S. Department of Energy Office of Science User Facility operated under the same contract as above and the Computational HEP program in The Department of Energy’s Science Office of High Energy Physics provided resources through the “Cosmology Data Repository” project (Grant \#KA2401022).}
X.H. acknowledges the University of San Francisco Faculty Development Fund.
\aguu{Support for \hst program 15867 was provided by NASA through a grant from the Space Telescope Science Institute, which is operated by the Association of Universities for Research in Astronomy, Inc., under NASA contract NAS 5-26555.}
\agu{S.H.S.~thanks the Max Planck Society for support through the Max Planck Research Group. E.J. acknowledges funding from Excellence Initiative of Aix-Marseille University - A*MIDEX, a French ``Investissements d'Avenir'' program (AMX-19-IET-008 - IPhU). Y.S. acknowledges support from the Max Planck Society and the Alexander von Humboldt Foundation in the framework of the Max Planck-Humboldt Research Award endowed by the Federal Ministry of Education and Research. This research used resources of the National Energy Research Scientific Computing Center, a DOE Office of Science User Facility supported by the Office of Science of the U.S. Department of Energy under Contract No. DE-AC02-05CH11231 using NERSC award HEP-ERCAP0021270.}

\software{
    TensorFlow \citep{TensorFlow},
    TensorFlow Probability \citep{dillon2017a}, 
    JAX \citep{bradbury2018a}, 
    Optax \citep{optax2020},
    lenstronomy \citep{birrer2018a},
    emcee \citep{foreman2013a},
    Matplotlib \citep{hunter2007a},
    seaborn \citep{waskom2021a},
    corner.py \citep{foreman2016a},
    TinyTim \citep{krist2011a},
    NumPy \citep{harris2020a}
}

\bibliographystyle{aasjournal}
\bibliography{dustarchive}

\appendix

\restartappendixnumbering
\renewcommand{\thefigure}{A\arabic{figure}}
\renewcommand{\theHfigure}{A\arabic{figure}}

Below, we show modeling results for four types of typical systems: folds, cusps, crosses, and doubles. \begin{figure}[H]
    \centering
    \includegraphics[width=0.9\textwidth]{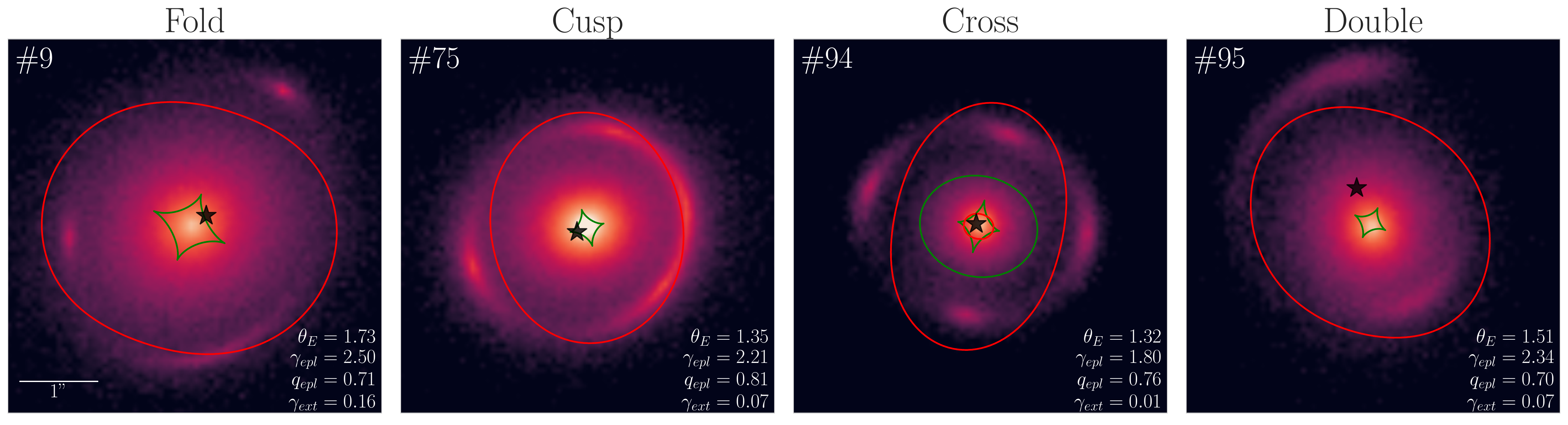}
    \caption{Four archetypal lensing systems selected from \cref{fig:lens-sample}: a folds, cusps, crosses, and doubles. The numbers in the top left corner refer to the ordering in \cref{fig:lens-sample}. All four systems have a comparable SNR. The source location is marked with a star, the critical curves are in red, and the caustics are in green. Observe the presence of an inner critical curve and caustic in the cross system, due to the fact that $\gamma_{epl} < 2$ \citep[e.g.,][]{oriordan2019}.}
    \label{fig:special-sys}
\end{figure}
For all four systems, the posterior mean agrees (within uncertainty) with the ground truth. In the folds (a), we point out the clear banana-shaped posterior (for similar examples with cluster lensing, see \citet{jullo2007}), as well as the weaker constraint on $\gamma$ (the standard deviation is here $\pm 0.1$, compared to the more typical $\pm 0.05$ for the other systems). This is worth keeping in mind when doing density profile slope studies. Furthermore, note the tendency for the VI posterior in (a) to underestimate the posterior scale. In contrast, for the cusp (b), the VI posterior overestimates the posterior scale for $\theta_E$. Notably, for the cross system (c), the results are qualitatively similar to the results for the reference system (which is also an approximate cross). That is, the degree of agreement between the ground truth and posterior mean is comparable to that of the reference system, and in both cases, the marginals of the VI posterior are similar to those of the true posterior. Finally, for the double (d), the marginals of the VI nearly perfectly agree with those of the true posterior.
\begin{figure}[H] 
\gridline{\fig{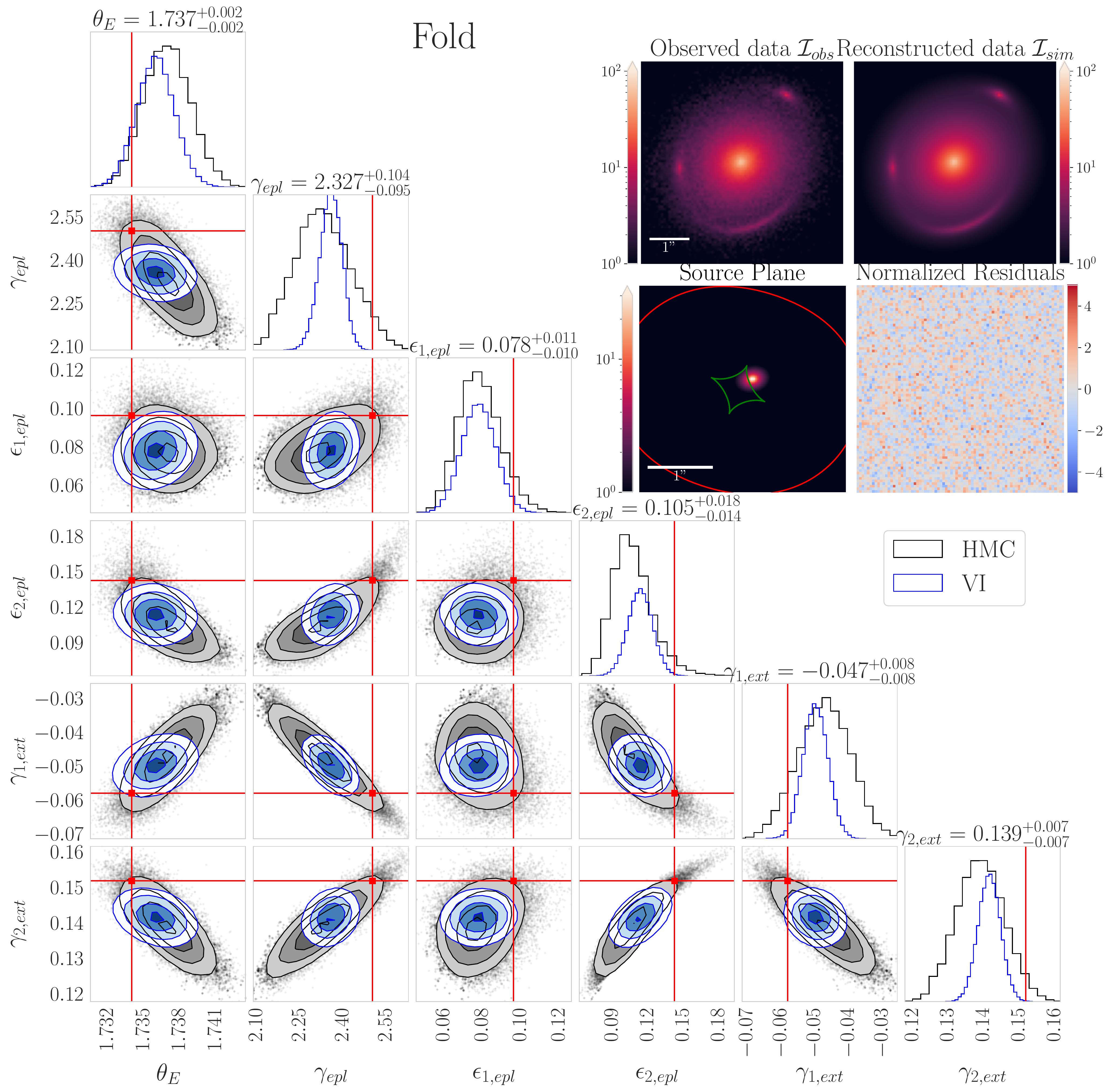}{0.49\textwidth}{(a) Folds}
          \fig{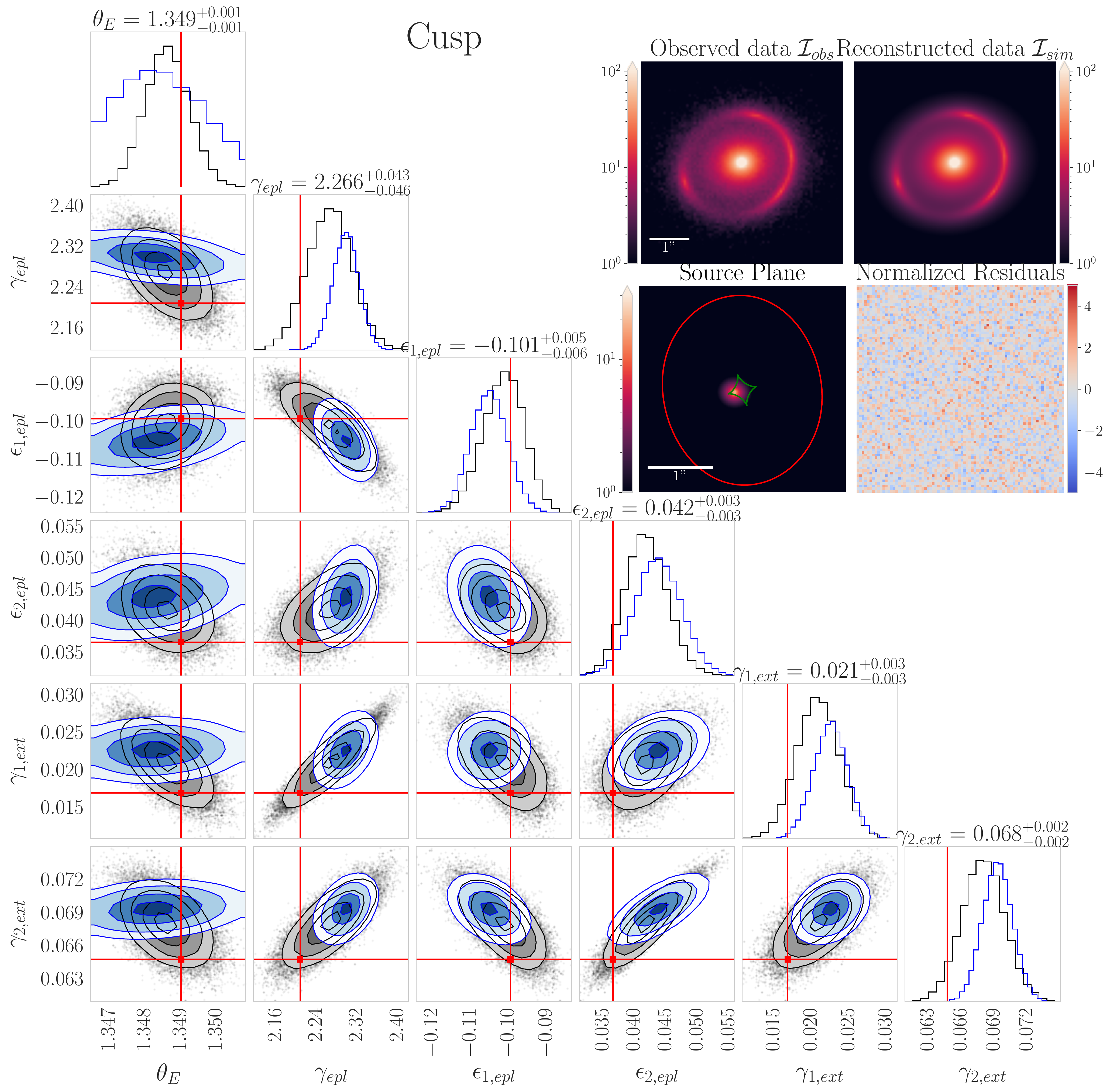}{0.49\textwidth}{(b) Cusps}}
\gridline{\fig{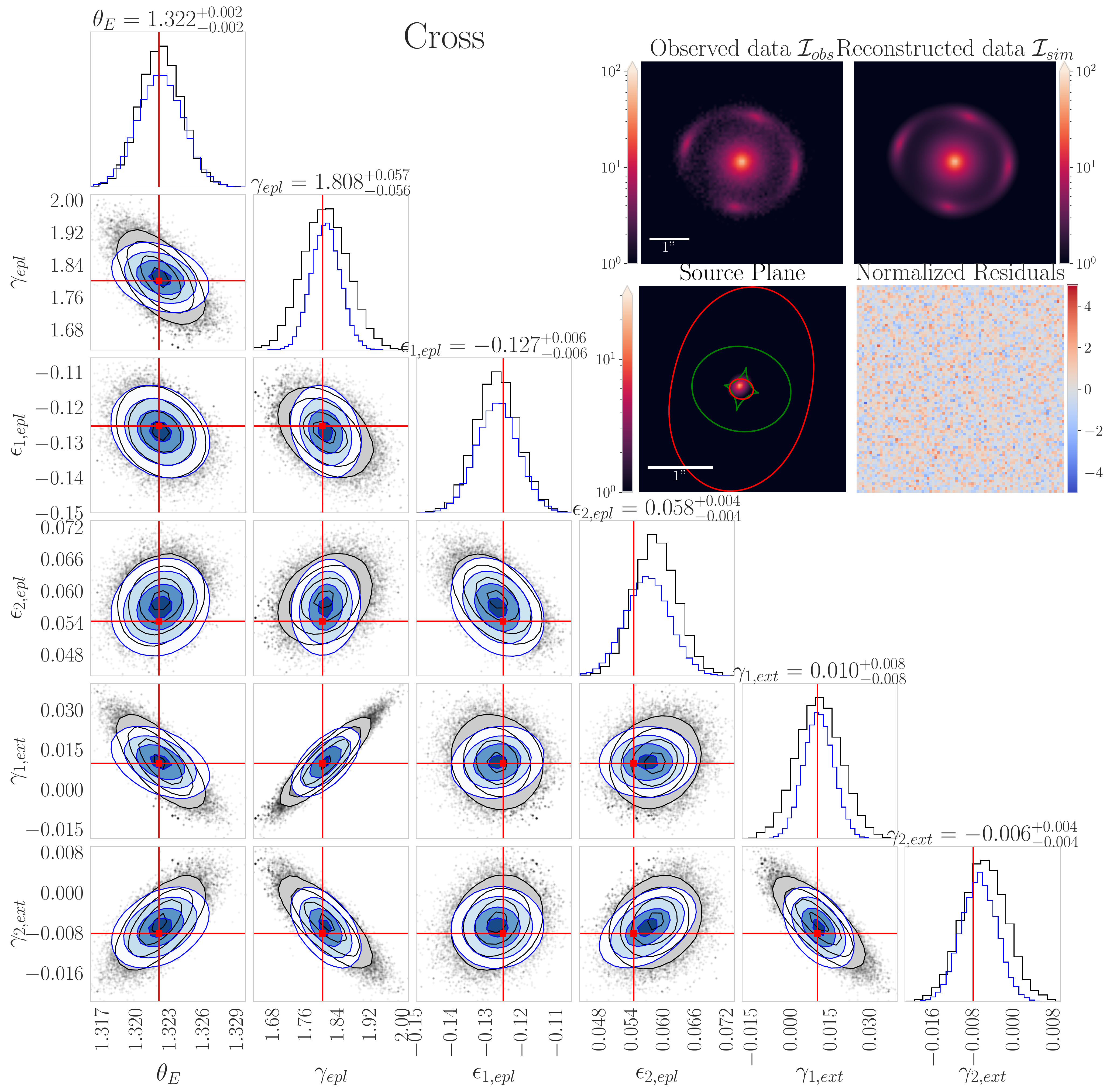}{0.49\textwidth}{(c) Crosses}
          \fig{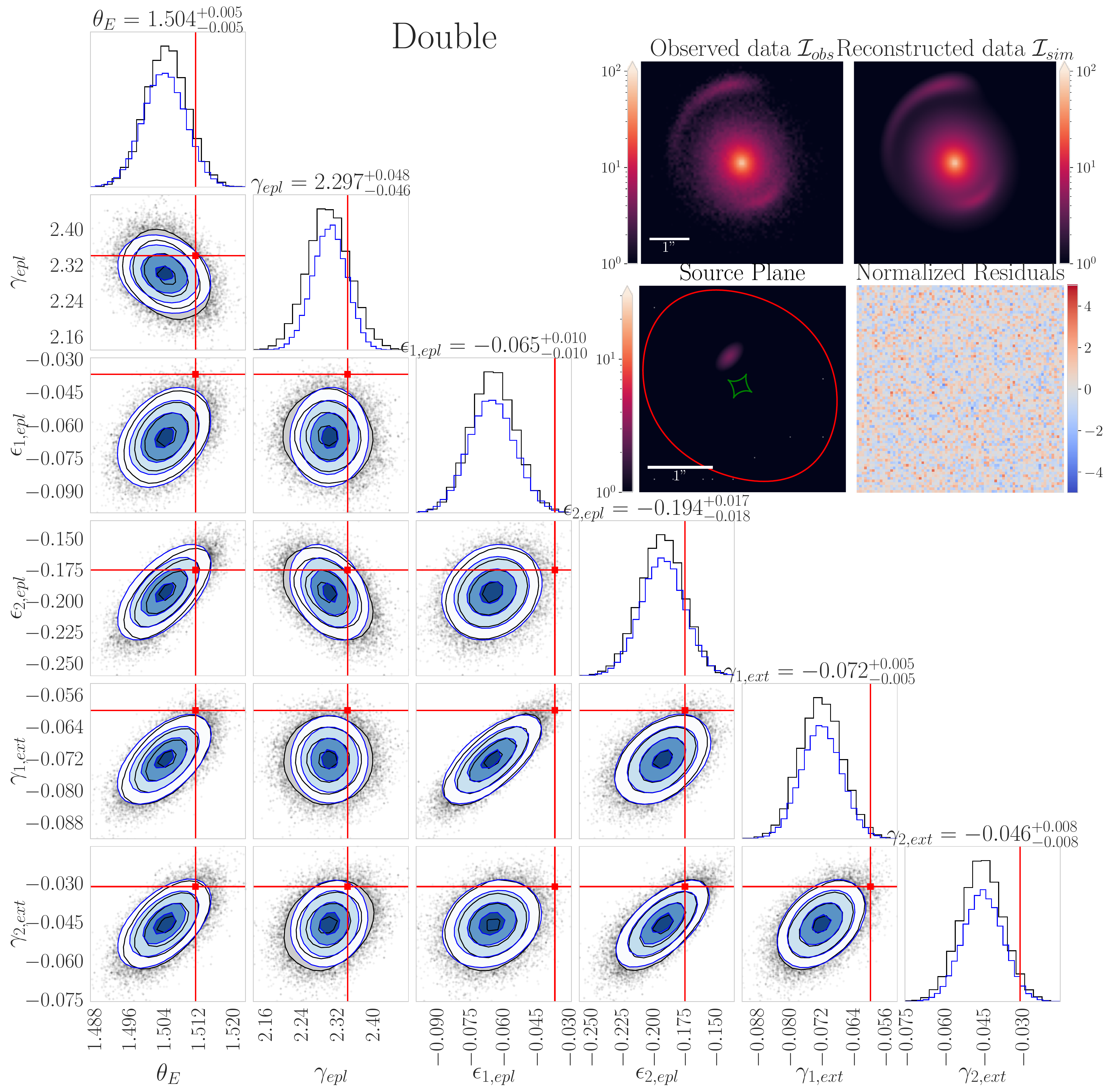}{0.49\textwidth}{(d) Doubles}}
\caption{Modeling results for each of the four archetypal systems. The samples and $0.5$, $1$, $1.5$, $2$ $\sigma$ contours (corresponding to roughly $12\%$, $39\%$, $68\%$, $86\%$ of the probability mass), for both VI and HMC are shown in blue and grey, respectively. \label{fig:archtypes}}
\end{figure}




\end{document}